\documentclass[%
 reprint,%
 amssymb, amsmath,%
 aip,cha,%
]{revtex4-1}

\usepackage{graphicx}
\graphicspath{ {images/} }
\usepackage{docs}%
\usepackage{bm}%
\usepackage{soul}
\usepackage[colorlinks=true,linkcolor=blue]{hyperref}%
\expandafter\ifx\csname package@font\endcsname\relax\else
 \expandafter\expandafter
 \expandafter\usepackage
 \expandafter\expandafter
 \expandafter{\csname package@font\endcsname}%
\fi
\begin{document}

\title{Measurement of turbulent spatial structure and kinetic energy spectrum by exact temporal-to-spatial mapping}%


\author{Preben Buchhave}
 \email{buchhavepreben@gmail.com}
\affiliation{Intarsia Optics, S{\o}nderskovvej 3, 3460 Birker{\o}d, Denmark}

\author{Clara M. Velte}%
 \email{cmve@dtu.dk}
 \affiliation{Department of Mechanical Engineering, Technical University of Denmark, \\ Nils Koppels All\'{e}, Bldg. 403, 2800 Kgs. Lyngby, Denmark}

\date{May 2017}%
\revised{August 2017}%

\begin{abstract}
We present a method for converting a time record of turbulent velocity measured at a point in a flow to a spatial velocity record consisting of consecutive convection elements. The spatial record allows computation of dynamic statistical moments such as turbulent kinetic wavenumber spectra and spatial structure functions in a way that completely bypasses the need for Taylor's Hypothesis. The spatial statistics agree with the classical counterparts, such as the total kinetic energy spectrum, at least for spatial extents up to the Taylor microscale. The requirements for applying the method is access to the instantaneous velocity magnitude, in addition to the desired flow quantity, and a high temporal resolution in comparison to the relevant time scales of the flow. We map, without distortion and bias, notoriously difficult developing turbulent high intensity flows using three main aspects that distinguish these measurements from previous work in the field; 1) The measurements are conducted using laser Doppler anemometry and are therefore \textit{not} contaminated by directional ambiguity (in contrast to, e.g., frequently employed hot-wire anemometers); 2) The measurement data are extracted using a correctly and transparently functioning processor and is analysed using methods derived from first principles to provide \textit{unbiased} estimates of the velocity statistics; 3) The exact mapping proposed herein has been applied to the high turbulence intensity flows investigated to avoid the significant distortions caused by Taylor's Hypothesis. The method is first confirmed to produce the correct statistics using computer simulations and later applied to measurements in some of the most difficult regions of a round turbulent jet -- the non-equilibrium developing region and the outermost parts of the developed jet. The proposed mapping is successfully validated using corresponding directly measured spatial statistics in the fully developed jet, even in the difficult outer regions of the jet where the average convection velocity is negligible and turbulence intensities increase dramatically. The measurements in the developing region reveal interesting features of an incomplete Richardson-Kolmogorov cascade under development.
\end{abstract}

\maketitle


\section{\label{sec:intro}Introduction}

Measurements of spatial structures in high Reynolds number turbulent flows are important for development and verification of turbulence models and indeed for the understanding of fundamental properties of turbulence. Essential to this problem is the measurement of high resolution statistical spatial quantities such as moments and turbulent kinetic energy spectra. Since the overwhelming number of high resolution measurements are obtained as time records with a probe located at a fixed point in space (e.g. hot-wire anemometer, HWA, and laser Doppler anemometer, LDA), a recurrent problem has been the conversion of time records into spatial records.

Taylor's hypothesis (TH) has been an invaluable method in turbulence research since Taylor presented the idea in 1938. In his seminal paper~\cite{Taylor1938}, Taylor proposed that the spatial fine-scale turbulent velocity structure is transported by the local mean velocity so quickly that the small scales do not have time to change, ``so that an unchanging pattern of turbulent motion is swept past a stationary probe'' (e.g. a hot wire probe). It is then possible to define a spatial sampling interval $ds_{TH}$ by the relation $ds_{TH}=\overline{u_1(\mathbf{x}_0,t)}\,dt$, where $\mathbf{x}_0$ is the location of the fixed measurement point (MP), $\overline{u_1(\mathbf{x}_0,t)}$ is the mean velocity in the average flow direction at the MP and $dt$ is the time increment. The temporal record, $t \in [0,T]$, is mapped into a spatial record $s \in [0,L]$ by the linear transformation:
\begin{equation}\label{eq:1}
s_{TH}(t)=\int_{t'=0}^t\overline{u_1(\mathbf{x}_0,t')}\,dt'.
\end{equation}
Under this condition of ``frozen turbulence'', the measured temporal record is interpreted (mapped) as a spatial homogeneous record upstream from the MP. In order for the statistical quantities to be valid, the method requires local homogeneity along the upstream mean flow direction.

However, it soon became clear that TH is not adequate in highly turbulent flows. Lin~\cite{Lin1953} was the first to evaluate TH in a shear flow, and further investigations of its limitations were presented in subsequent work~\cite{FisherDavies1964,Lumley1965,WnC1977,Gurvich1980,Antonia1980,DeardorffWillis1982,Kaimal1982}.

Heskestad~\cite{Heskestad1965} proposed a generalized form of TH in which the convection due to the large fluctuating velocity components was taken into account. Lumley~\cite{Lumley1965} further examined the magnitude of the various terms of a generalized TH, which takes into account the fact that all spatial structures are carried past the probe by larger eddies, and that these eddies form a continuum of scales. He concluded that the most important effect on the temporal energy\footnote{The temporal energy spectrum is often in the turbulence community refereed to as the power spectrum (of the velocity fluctuations) by convention from electrical engineering.} spectrum derives from the convection due to the large energy carrying eddies, the separation between ``large eddies'' and ``small isotropic eddies'' chosen somewhat arbitrarily (but later confirmed more rigorously) as $k_1 / 2\pi \gg u_1' / \overline{u}_1$, where $k_1$ is the wave number, $u_1'$ and $\overline{u}_1$ are the spatial velocity gradient and the mean velocity in the flow direction, respectively. Lumley further introduced a correction to the spatial spectrum derived by expanding the characteristic function and keeping only terms to second order and argued that this correction for all practical purposes would not exceed approximately 30\%.

Over the ensuing years, numerous studies investigated TH by comparison to measurements. For example, Wyngaard and Clifford~\cite{WnC1977} tested TH by comparing measured atmospheric data to a Gaussian fluctuating convection velocity model and to Lumley's correction. Further important studies of the relation between temporal and spatial scales were published by e.g. Antonia \textit{et al.}~\cite{Antonia1980}, Tennekes~\cite{Tennekes1975}, Thacker~\cite{Thacker1977}, Champagne~\cite{Champagne1978}, Zaman and Hussain~\cite{ZamanHussain1981} and Mi and Antonia~\cite{MiAntonia1994}, to mention just a few.

More recent studies of the errors committed by applying TH to highly fluctuating velocities as in atmospheric turbulence were published by E. Gledzer~\cite{Gledzer1997}, who studied the shape of the spatial energy spectrum resulting from different correction methods and the exclusion of the acceleration terms in Navier-Stokes equations. M. Wilczek and co-workers~\cite{Wilczek2014} also studied the effect of large-scale random sweeping velocities on the determination of the Kolmogorov constant.

Our method relies on the simultaneous measurement of the desired flow quantity and the magnitude of the instantaneous velocity vector. \textit{We propose that the instantaneous velocity magnitude is the relevant quantity that transports the fluid properties such as small scale velocity structures as well as scalar quantities such as e.g. temperature and particle concentration past a stationary probe and thus should be the quantity relevant for a mapping of temporal records into spatial ones.} The idea seems to have occurred to just a few researchers in the past, possibly because of the perceived difficulties such as simultaneous 3D measurements of velocity components and the resulting irregular spatial sampling intervals that preclude the use of the fast Fourier transform. R. J. Hill~\cite{Hill1996} considers a 3D measurement of the short-time averaged mean velocity to obtain the velocity in a coordinate system turning with the instantaneous flow direction from which the velocities in the lab-coordinates can be obtained. However, he dismisses the idea because it involves a continually changing coordinate transformation and a non-equidistant sampling scheme. Pinton and Labb\'{e}~\cite{PnL1994} use the averaged velocity over one revolution in a swirling flow as the basis for applying TH. This is an improvement over a long time average, but does not account for the randomly fluctuating convection velocity during a single revolution.

In the following, we introduce the true temporal-to-spatial mapping and discuss statistical quantities evaluated with the new method, in particular first order static moments and second order dynamic moments, and compare the results to moments computed by time averaging. In subsequent sections we describe the method applied to HWA and LDA measurements. We continue by discussing the range of equivalence between spectra measured by our method and by conventional methods. One of the results of this analysis is that it is important to consider if an instrument performs inherently ``temporal sampling'' or inherently ``spatial sampling''. For example, we shall show that a digitally sampled HWA performs temporal sampling, whereas an LDA performs spatial sampling. The type of sampling has consequences for interpretation and computation of static and dynamic moments. In Section III, we illustrate the validity of the concept by showing that the new method can restore simulated spatial records while Taylor's Hypothesis is not able to do so.

In the last section, we verify experimentally our method by comparing spatial energy spectra from LDA measurements in the developed jet to corresponding spatial spectra measured along mapped homogeneous directions using particle image velocimetry (PIV). The agreement between these spectra even in the outer parts of the jet is then contrasted to the spatial spectra obtained using Taylor's hypothesis. We then apply the method to LDA measurements in the developing, non-equilibrium part of a round, turbulent jet in air and compare spatial energy spectra and spatial 2nd order structure functions measured in that region to the same quantities measured further downstream in the fully developed part of the jet. We discuss the shape of these functions and interpret them as due to incomplete, not fully developed Richardson-Kolmogorov cascades.

\section{\label{sec:theory}Theory}

\begin{figure}[b]
\includegraphics[width=0.45\textwidth]{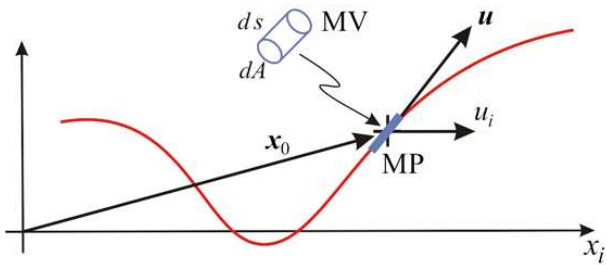}
\caption{\label{fig:taylor2_figure_1_a} The instantaneous convection element through the MV.}
\end{figure}

We consider an experiment in which we measure some property of the fluid at a fixed point in space, $\mathbf{x}_0$, the measurement point (MP, see Figure~\ref{fig:taylor2_figure_1_a}). The property in question could typically be a component, $u_i=u_i(\mathbf{x}_0, t)$, say, of the three-dimensional velocity vector $\mathbf{u}(\mathbf{x}_0, t)$ recorded at $\mathbf{x}_0$ as a continuous function of time, $t$. $\mathit{\mathbf{i}}=\{1,2,3\}$ indicates the three orthogonal coordinate axes. Essential to the method is, that in addition to the quantity of interest, we measure the magnitude of the instantaneous velocity vector, $u(t) \equiv u(\mathbf{x}_0,t) = |\mathbf{u}(\mathbf{x}_0,t)|$. How this is done will depend on the actual experiment, whether it is for example a hot-wire anemometer or a laser Doppler anemometer or perhaps some other method.

During the infinitesimal time element $dt$, the fluid velocity $\mathit{\mathbf{u}}$ transports a fluid element $dV = dA \cdot ds$ through the infinitesimal measurement volume (MV), where $dA$ is the cross section and $ds = u\,dt$ is the length of the infinitesimal record convected through the measuring volume. This finite length element $ds$ represents a piece of a spatial record of the fluid having passed through the MV with its associated physical properties. By adding these consecutive convection elements, we form a spatial record, $s$, which we have termed the \textit{convection record}:
\begin{equation}\label{eq:2}
s(t) = \int_{0}^t u(\mathbf{x}_0,t')\,dt'
\end{equation}
Although $s$ is the scalar length of the accumulated convection elements for the fluid passing through the MP, and although $s$ has the dimension of length, the measurement is still basically a temporal measurement where we have converted the temporal record to a spatial one according to Equation~(\ref{eq:2}). We may describe the procedure as an improved version of TH, where instead of using the local mean velocity to convert from the temporal to the spatial domain according to the formula  $ds_{TH} = |\overline{u_1 (\mathbf{x}_0,t)}|dt$, we use the instantaneous velocity magnitude $u(t) = |\mathbf{u}(\mathbf{x}_0,t)|$, measured together with the desired physical property. As explained below, this represents a mapping from a record indicating the observation time to a record indicating the volume flow density through the MP. Note that the only requirement for using this method to obtain statistical quantities is that the flow is stationary at the measurement point.

\subsection{\label{sec:staticmoments}Static moments}

We can now compute the moments of any measurable physical quantities, e.g. the velocity component $u_i \equiv u_i(s)$, recorded as a function of $s$. Let us take as a generic example the first moment:
\begin{equation}\label{eq:3}
\langle u_i \rangle_s = \lim_{L \rightarrow \infty} \frac{1}{L}\int_{0}^L u_i(s)\,ds
\end{equation}
where we shall reserve the bracket $\langle \, \rangle_s$ for moments using the spatial record $s$. $L$ is here the length of a finite spatial record, $L = \int_0^T u (\mathbf{x}_0,t)\, dt$, where $T$ is the length of the corresponding temporal record. The question is how this spatial mean relates to the temporal mean
\begin{equation}\label{eq:4}
\overline{u}_i = \lim_{T \rightarrow \infty} \frac{1}{T}\int_{0}^T u_i(t)\,dt
\end{equation}
Since the mapping from the temporal to the spatial domain represented by Equation~(\ref{eq:2}) is a nonlinear one, the two moments cannot be identical. Recalling that $s$ represents the length of the accumulated convection elements or convection record through the infinitesimal MV and that $dA\, u \, dt = dA \, ds$ is a volume element, we conclude that whereas the moment $\overline{u}_i $ represents the conventional temporal mean velocity of the component $u_i(t)$ at the MP, the moment $\langle u_i \rangle_s$, represents the average volume flow density along the \textit{i}-axis at the MP.

\subsection{\label{sec:dynamicmoments}Second order dynamic moments}

Most measurements with low noise and high dynamic spectral range are time measurements, predominantly HWA measurements, and these measurements register the fluctuations of the time signal as the fluid passes the stationary probe. However, the temporal energy spectrum does not display the correct distribution of the spatial scales in the turbulent flow due to the convection effect of the large scale velocity fluctuations that tend to sweep the small scales past the probe with varying velocity.

TH does not compensate for the mapping effect of the large velocity fluctuations, and especially at high levels of turbulence the form of the spatial spectrum may be quite different from that of the temporal spectrum~\cite{PBCargese2016}. By applying our method and using the measured velocity magnitude to perform the time-to-space conversion as indicated in Equation~(\ref{eq:2}), we obtain a correct representation of the small scale spatial structures.

We shall consider two forms of the converted one-dimensional spatial autocovariance function and spatial energy spectrum according to the quantity we choose to measure:

A single velocity component, $u_i(s)$:
\begin{equation}\label{eq:5}
C_{u_i}(r) = \langle u_i(s)u_i(s+r) \rangle
\end{equation}
where the brackets indicate ensemble mean over many realizations.

Also of interest is the autocovariance function of velocity magnitude, $u(s)$:
\begin{equation}\label{eq:6}
C_{u}(r) = \langle u(s)u(s+r) \rangle
\end{equation}
The corresponding spatial energy spectra are given by
\begin{equation}\label{eq:7}
F_{u_i}(k_i) = \int_{-\infty}^{\infty} e^{-i2\pi k_i r}C_{u_i}(r)\,dr
\end{equation}
and
\begin{equation}\label{eq:8}
F_{u}(k) = \int_{-\infty}^{\infty} e^{-i2\pi k r}C_{u}(r)\,dr
\end{equation}
or by the direct method
\begin{equation}\label{eq:9}
F_{u_i}(k_i) = \frac{1}{L}\tilde{u}_i(k_i) \tilde{u}_i(k_i)^{\ast}, \quad F_{u}(k) = \frac{1}{L}\tilde{u}(k) \tilde{u}(k)^{\ast}
\end{equation}
where
\begin{equation}\label{eq:10}
\tilde{u}_i(k_i) = \int\limits_0^L\ e^{-i 2 \pi k_i s}u_i(s)\, ds
\end{equation}
and
\begin{equation}\label{eq:11}
\tilde{u}(k) = \int\limits_0^L\ e^{-i 2 \pi k s}u(s)\, ds
\end{equation}
are the Fourier transforms performed over a finite length $L$ of the convection record.

\section{\label{sec:HWA}Temporal sampling -- the HWA case}

\subsection{Sampling}

Hot-wire anemometers deliver analog time signals that are usually digitized with regular sampling intervals $\Delta t = 1/\nu$ where $\nu$ is the sampling frequency. The sampled velocity $u_0(t)$ is then
\begin{equation}\label{eq:12}
u_0(t) = \frac{1}{\nu}\sum_{n=1}^N \delta (t-t_n) u(t)
\end{equation}
and the individual samples are given by
\begin{equation}\label{eq:12b}
u_n = \delta (t-t_n) u(t) =u(t_n)
\end{equation}
where $N$ is the number of samples in a record. In practice, the instantaneous velocity magnitude must be obtained by a separate measurement, either by a 3-D HW probe or by a separate measurement by an omnidirectional small probe located near the wire measuring the relevant velocity component.

Estimates for the static moments can then be found by arithmetic averaging over the record, for example mean velocity:
\begin{equation}\label{eq:13}
\widehat{\overline{u}}_0 = \frac{1}{T}\int_{0}^T \frac{1}{\nu} \sum_{n=1}^N \delta (t-t_n) u(t)\, dt = \frac{1}{N} \sum_{n=1}^N u_n
\end{equation}
where the hat indicates that we are dealing with an estimate based on one record.

The temporal second order dynamic moments of course represent the energy of the velocity signal detected by the probe. However, due to the ``harmonium effect'', the frequency shift due to the convection of the small spatial scales by the large convecting eddies, the convected dynamic moments do not represent the spatial velocity structure or the energy content in the velocity pattern. The separation in convected and convecting eddies is actually somewhat artificial; the frequency shift is due to the total magnitude of the velocity vector, but the most significant effect is due to the large scales~\cite{Lumley1965}. To get an unbiased measurement of the small spatial structures we need to convert from the temporal record $t_n$ to a spatial one $s_n$:
\begin{equation}\label{eq:14}
\Delta s_n = u_n\Delta t, \quad s_n = \sum_{n'=1}^n \Delta s_{n'} = \sum_{n'=1}^n \Delta u_{n'} \Delta t
\end{equation}

In the spatial domain, the very same samples are used even if their distribution is different along the spatial and temporal records. Therefore, the temporal mean velocity in the spatial domain is still given by the arithmetic average over the measured samples
\begin{equation}\label{eq:15}
\langle u_0 \rangle = \frac{1}{N} \sum_{n=1}^N u_n
\end{equation}
The spatial mean needs to be corrected for sampling bias in the spatial domain. This could be done by resampling the spatial record with equal sampling increments. 

We can now compute the spatial energy spectra:
\begin{equation}\label{eq:16}
\widehat{F}_{u_i}(k_i) = \frac{1}{L}\tilde{u}_i(k_i) \tilde{u}_i(k_i)^{\ast}, \quad \widehat{F}_{u}(k) = \frac{1}{L}\tilde{u}(k) \tilde{u}(k)^{\ast}
\end{equation}
where the tilde indicates Fourier transform computed over the measured samples and the hat indicates an estimate based on a single record. However, the samples are not equidistantly spaced on the spatial record, so the Fourier transform must be computed by the discrete Fourier transform, DFT:
\begin{equation}\label{eq:17}
\tilde{u}_i(k_i) = \sum_{n=1}^N e^{-i 2 \pi k_i s_n} u_{i,n}\Delta s_n
\end{equation}
and
\begin{equation}\label{eq:18}
\tilde{u}(k) = \sum_{n=1}^N e^{-i 2 \pi k s_n} u_{n}\Delta s_n
\end{equation}
As $\tilde{u}(k)$ is the Fourier transform of the velocity vector magnitude, $u$, which is always in the direction of the instantaneous velocity vector, the second form of the spectra in Eq.~(\ref{eq:16}) represents the total turbulent kinetic energy of the small scales.

\subsection{\label{sec:CGHWA}Computer generated HWA spectra}

To show that the method produces the desired statistics, we generate the HWA data as follows:

First we generate a large, slowly fluctuating Gaussian 3D velocity, $\mathbf{u}_l(t)$. Then we simulate a smaller spatially isotropic velocity with a von K\'{a}rm\'{a}n spectrum, $\mathbf{u}_s(s)$. The von K\'{a}rm\'{a}n spectrum is chosen with an exponential roll-off that models as closely as possible the jet spectrum described in the experiment below:
\begin{equation}\label{eq:19}
S_{vK}(k) = \frac{1}{62.5}\cdot\frac{1}{\left [ 1 + (k/45)^2 \right ]^{5/6}} \cdot \exp \left [ -(k/2500)^{4/3}\right ]
\end{equation}
This spatial spectrum is shown as the yellow curve in the following figures. The mean velocity was $1\, ms^{-1}$ and the record length was $1\, s$, which allows us to use the same scale on the abscissa for both temporal and spatial spectra.

\begin{figure*}
\begin{minipage}[b]{0.45\linewidth}
\centering
\includegraphics[width=1.0\textwidth]{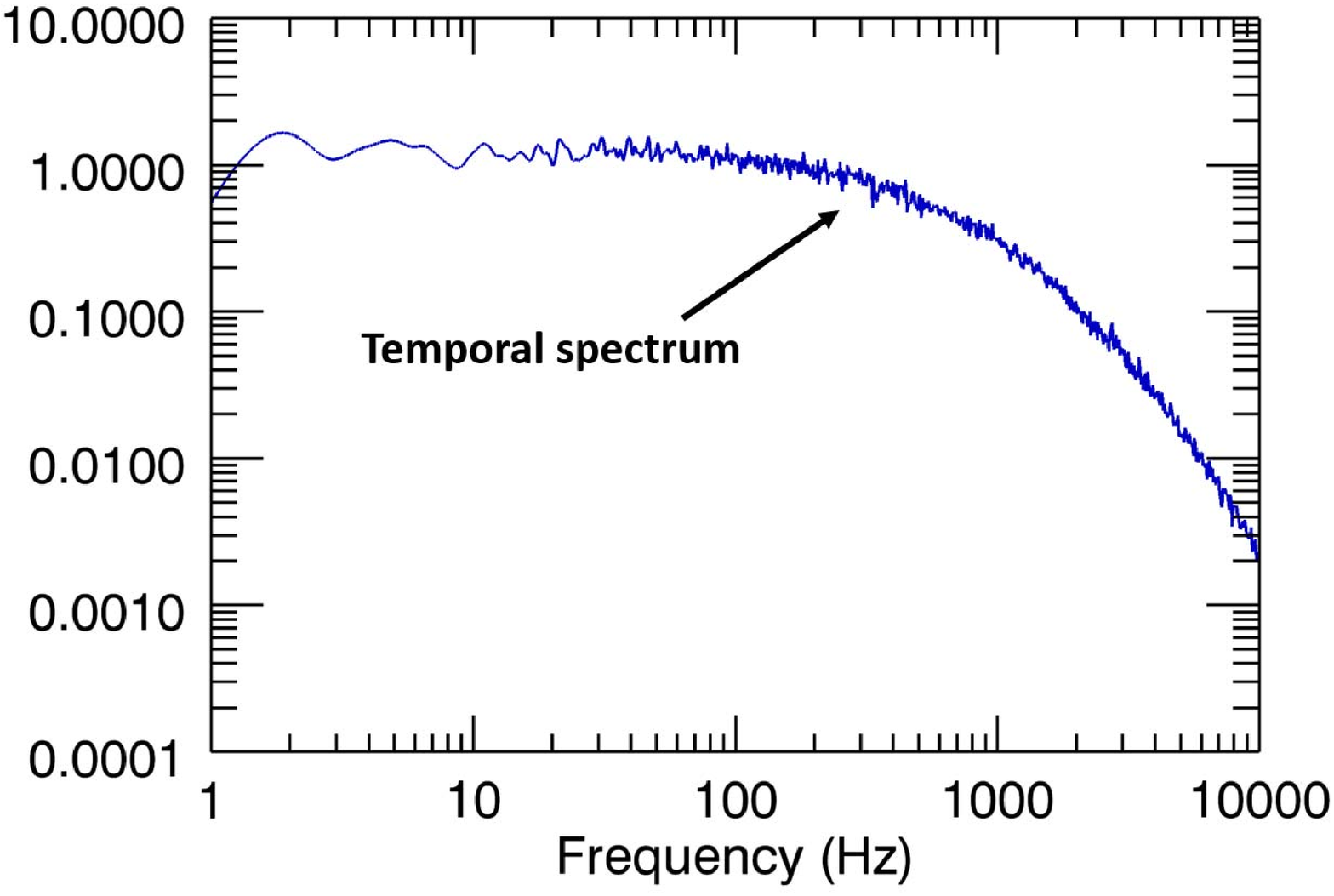}
\end{minipage}
\hspace{0.5cm}
\begin{minipage}[b]{0.45\linewidth}
\centering
\includegraphics[width=1.0\textwidth]{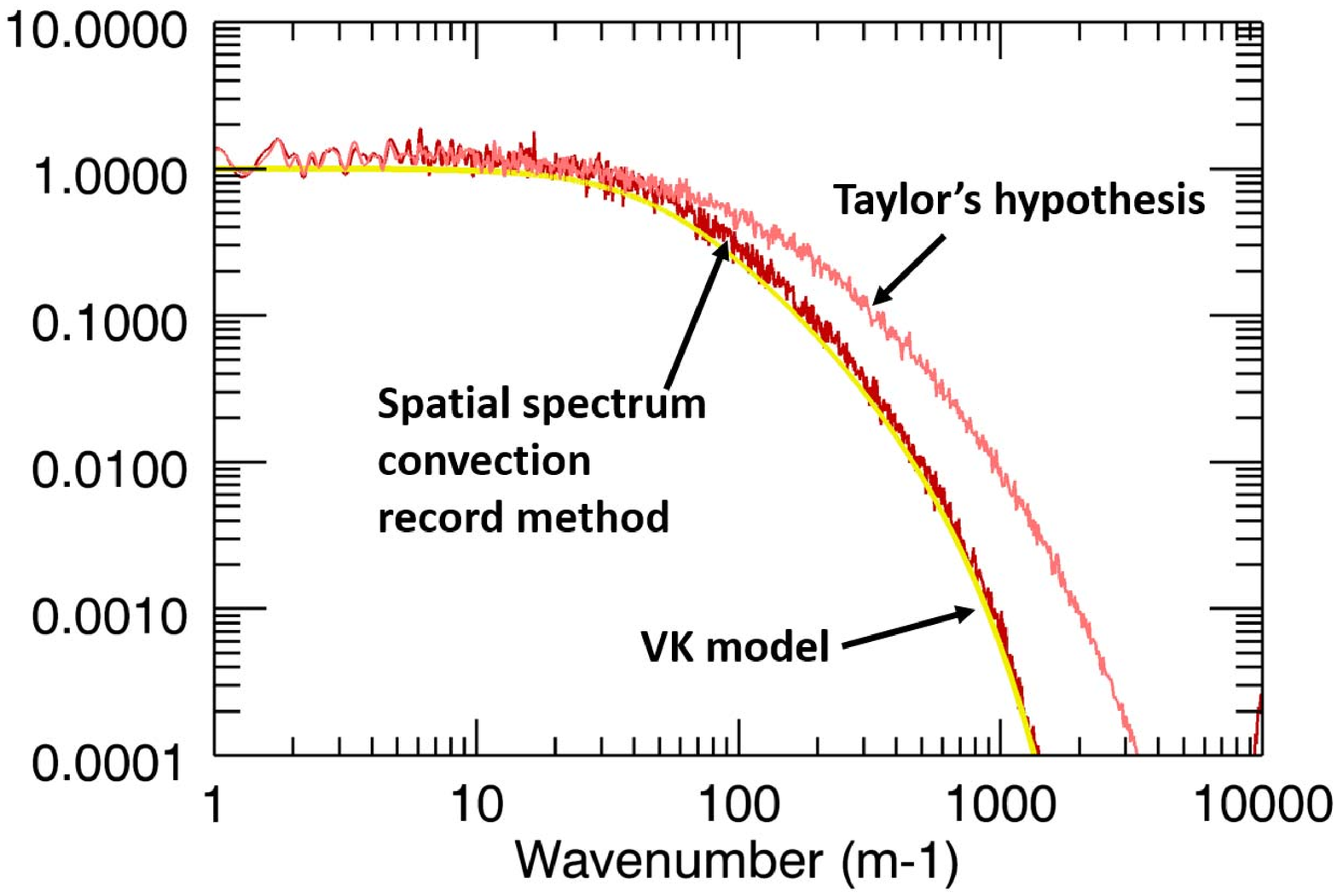}
\end{minipage}
\caption{\label{fig:Taylor-2_PR_fig2} Computer generated HWA spectra. Block average of 100 records. Yellow: VK-based model. Blue: temporal spectra. Red: spatial spectra. LHS: Temporal spectrum of resampled time record. RHS: Spatial spectrum by the new method (dark red) and converted by Taylor's method (light red). Color online. }
\end{figure*}

The total signal is assumed to be homogeneous and a continuous function of spatial coordinates (which we achieve with a sufficiently small primary spatial sampling interval, $\Delta s$). We use the slowly varying signal, which we assume to be the so-called energy containing, convecting velocity to transform the spatial record (that of the yellow spatial spectrum) to a temporal record measured at the MP according to the formula
\begin{equation}\label{eq:20}
\Delta t_n = \Delta s/u_{l,n}
\end{equation}
where $u_{l,n}$ is the magnitude of the n'th sample of the slowly varying (large scale) convection velocity. We then create a HWA-like signal by resampling the nonlinear record with a constant sampling interval, $\Delta t_{rs}$. The blue curve in Figure~\ref{fig:Taylor-2_PR_fig2} left represents the temporal energy spectrum based on this temporal record. We then apply both our new method and TH to convert to spatial records and obtain the spatial energy spectra as described above, see Figure~\ref{fig:Taylor-2_PR_fig2} right.

The turbulence intensity of the total velocity magnitude for these plots was 54\%, and it is clearly seen that the traditional Taylor's hypothesis is not valid in this case whereas the new method is able to restore the original spatial spectrum.

\section{\label{sec:SpatialSampling}Spatial sampling -- the LDA case}

\subsection{\label{sec:LDAsampling}Sampling}

The burst-type LDA is an example of an instrument performing spatial sampling or sampling in the spatial domain, because the sampling process is determined by the (assumed) uniform, albeit random, location of particles in the fluid. The sampled velocity in space is given by
\begin{equation}\label{eq:21}
u_0(s) = \frac{1}{\nu} \sum_{n=1}^N \delta (s-s_n) u(s)
\end{equation}
and the individual samples by
\begin{equation}\label{eq:21b}
u_n = \delta (s-s_n) u(s) = u(s_n)
\end{equation}

However, due to the random distribution of particles, the sampling intervals $\Delta s_n = s_n - s_{n-1}$ are not equal as in the case of the HWA. The temporal sampling seen by a stationary probe, e.g. a laser beam, will now be random with a mean rate of arrival proportional to the volume flow, assuming a constant MV cross section. Thus, in the LDA-case, both static and dynamic moments need to be corrected for velocity bias by residence time weighting in order to obtain correct time averages~\cite{Buchhave1979,BuchhaveGeorgeLumley1979,Velteetal2014a}.

To get the spatial spectrum, we map from the measured time record to the spatial domain by the formula
\begin{equation}\label{eq:22}
\Delta s_n = u_n \Delta t_n, \quad s_n = \sum_{n'=1}^n u_{n'} \Delta t_{n'}
\end{equation}
where $u_n$ is the sampled velocity magnitude. As the LDA measures components of the velocity, determination of the velocity magnitude requires a 3-D velocity measurement: $u_n = \sqrt{u_i^2 + u_j^2 + u_k^2}$. However, most modern LDA systems measure the so-called residence time or transit time for a particle traversing the MV. Knowing the diameter of the MV, $d_{MV}$, we can estimate $u_n$ as
\begin{equation}\label{eq:23}
u_n = d_{MV} /\Delta tr_n
\end{equation}
where $\Delta tr_n$ is the measured residence time. Due to the random path of particles through the measuring volume, the residence time will fluctuate around a mean value. Thus $u_n$ will be a strongly fluctuating quantity. However, as we compute the spectrum as an average (block average) over many spectral estimates, and since the particle path and the velocity are uncorrelated, the effect of the residence time fluctuations will be reduced. As we have shown previously in~\cite{Buchhaveetal2015}, the noise due to the random sampling is the dominant noise source, and the residence time fluctuations can be neglected. This is also confirmed by the convergence of the block averaged spectra shown below.

The time sampling intervals, $\Delta t_n$, are now different (random) and not known a priori. However, when the sampling rate is high enough, we can replace the convection  element by the time between samples multiplied by the latest measured velocity (or by some higher order interpolation scheme):
\begin{equation}\label{eq:24}
\Delta s_n = u_n (t_n - t_{n-1}), \quad s_n = \sum_{n'=1}^n u_{n'} (t_{n'} - t_{n'-1}).
\end{equation}
The spatial energy spectra can now be computed:
\begin{equation}\label{eq:25a}
\widehat{F}_{u_i}(k_i) = \frac{1}{L} \tilde{u}_i(k_i) \tilde{u}_i(k_i)^{\ast}
\end{equation}
and
\begin{equation}\label{eq:25b}
\widehat{F}_{u}(k) = \frac{1}{L} \tilde{u}(k) \tilde{u}(k)^{\ast}
\end{equation}
with
\begin{equation}\label{eq:26}
\tilde{u}_i(k_i) = \sum_{n=1}^N e^{-i 2 \pi k_i s_n} u_{i,n}\Delta s_n
\end{equation}
and
\begin{equation}\label{eq:27}
\tilde{u}(k) = \sum_{n=1}^N e^{-i 2 \pi k s_n} u_{n}\Delta s_n
\end{equation}

Again, as $u_n$ is the magnitude of the instantaneous velocity vector, Eqn.~(\ref{eq:25b}) represents the total kinetic energy spectrum.

\subsection{\label{sec:CGLDA}Computer generated LDA spectra}

The computer generated data are described in Buchhave \textit{et al.}~\cite{Buchhaveetal2014} and Velte \textit{et al.}~\cite{Velteetal2014b}. Briefly, the random spatial samples were grabbed from a high data rate primary velocity record with the von K\'{a}rm\'{a}n temporal energy spectrum Eq.~(\ref{eq:19}) by a Poisson process. Since we are sampling the velocity in space and assuming the particles to be uniformly distributed in space, the Poisson process is not modulated by velocity, and there is no velocity -- sample rate bias.

\begin{figure*}
\begin{minipage}[b]{0.45\linewidth}
\centering
\includegraphics[width=1.0\textwidth]{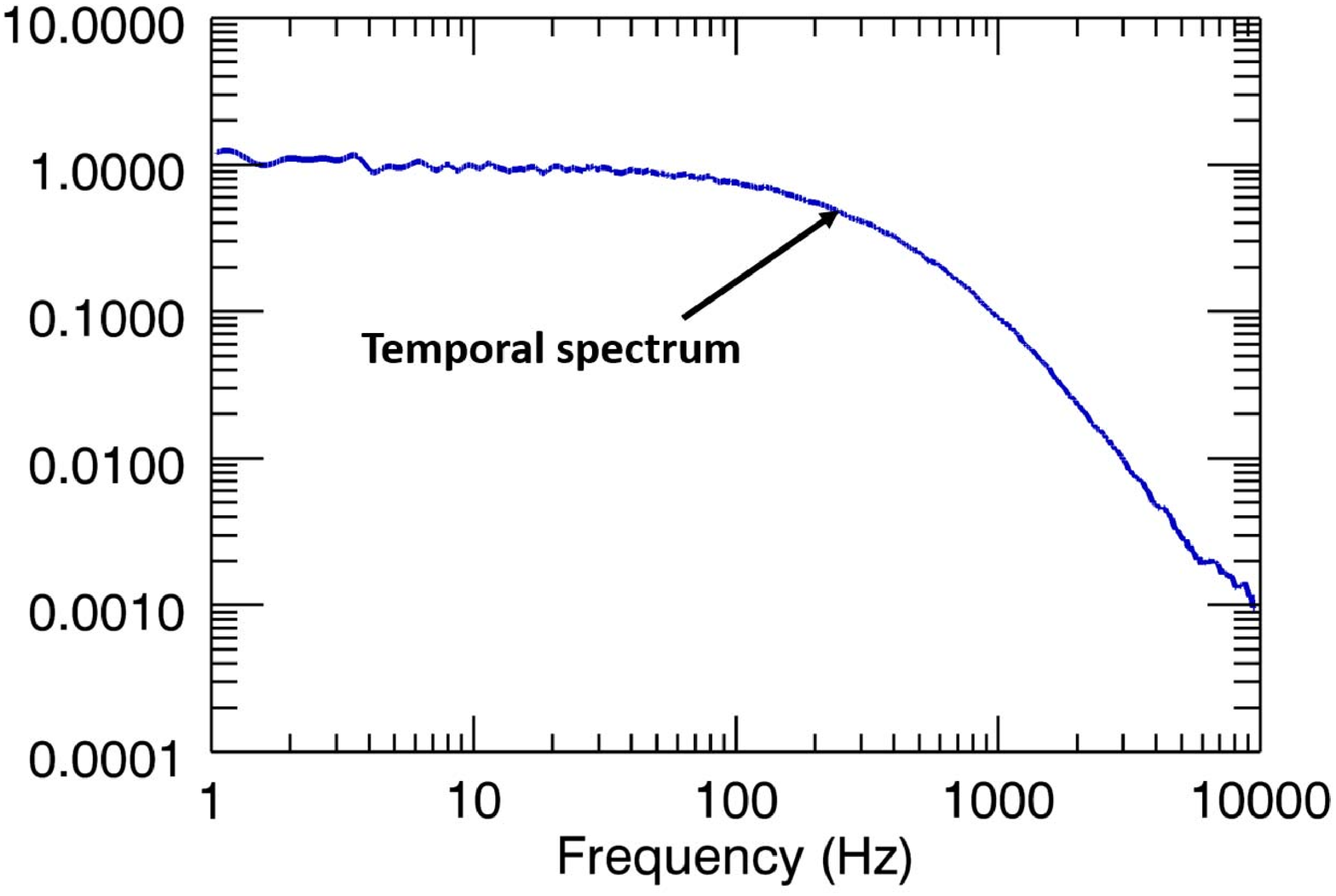}
\end{minipage}
\hspace{0.5cm}
\begin{minipage}[b]{0.45\linewidth}
\centering
\includegraphics[width=1.0\textwidth]{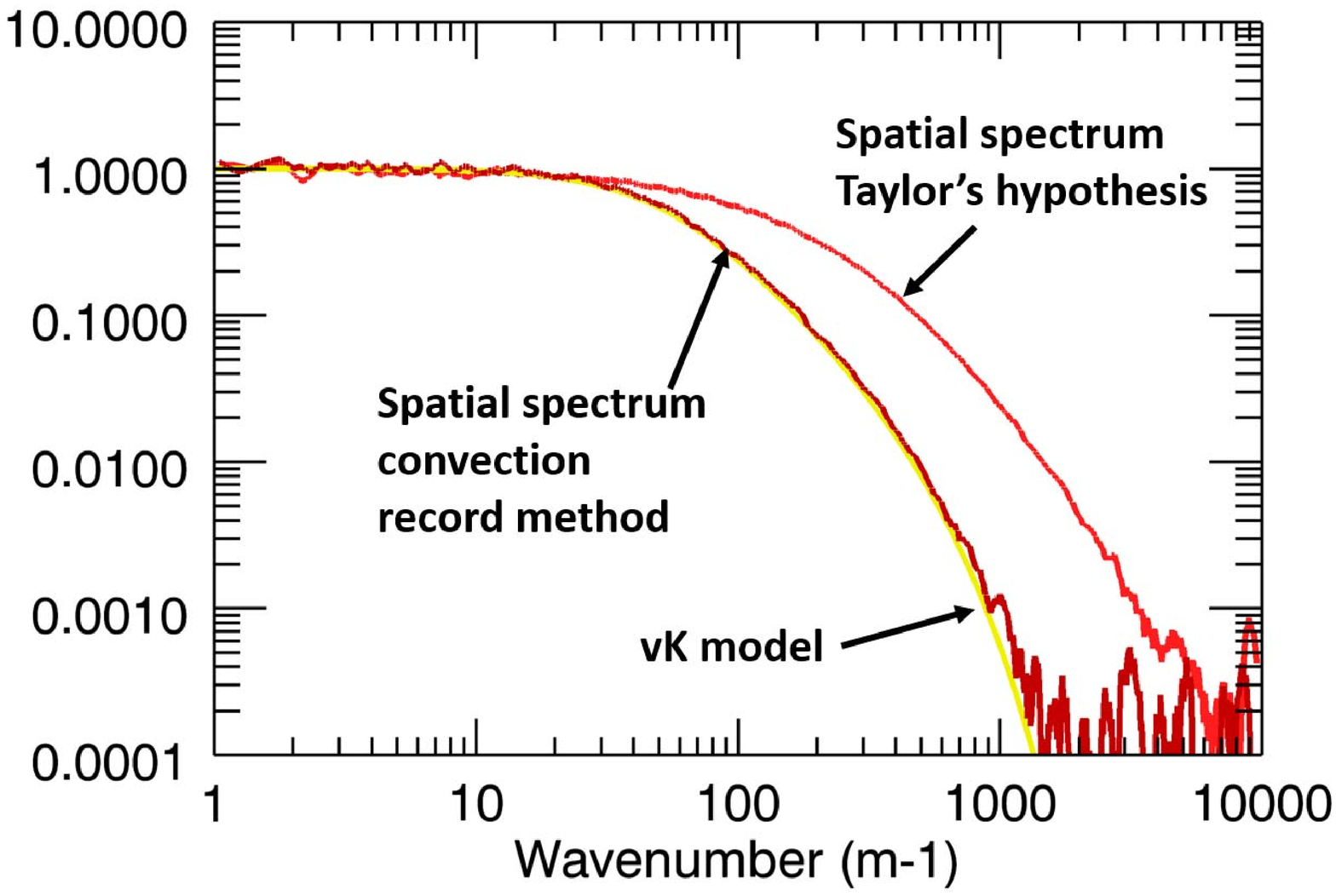}
\end{minipage}
\caption{\label{fig:Taylor2_PR_fig_3} Spectra of the von K\'{a}rm\'{a}n turbulence convected past the LDA MV by a large low frequency Gaussian fluctuation. Left: Blue: The temporal spectrum. Right: Yellow: The original von K\'{a}rm\'{a}n model spectrum. Dark red: The spatial spectrum restored. Light red: The spatial spectrum restored by the conventional Taylor's hypothesis. Color online. }
\end{figure*}

We now assume that the velocity fluctuations are convected through the LDA MV by 3-D Gaussian large scale eddies. The mean velocity was $1\, ms^{-1}$ and the record length was $1\,s$. The turbulence intensity for the complete fluctuating convection velocity signal was 54\%. The spectra were averaged over 100 records.

The resulting randomly sampled time record results in an aliased time spectrum (dark blue) as shown in Figure~\ref{fig:Taylor2_PR_fig_3} LHS. The RHS shows that the new method restores the von K\'{a}rm\'{a}n spectrum even in the presence of large scale low frequency 3-D fluctuations whereas Taylor's hypothesis in this case with a 54\% turbulence intensity does not adequately restore the spatial record.

\section{\label{sec:ClassicalSpectra}Relation to classical spectra}

In classical turbulence theory~\cite{Batchelor1953}, the starting point is a homogeneous, random velocity field observed at an arbitrary point in time. The 3-D covariance tensor is defined as
\begin{equation}\label{eq:28}
\mathbf{R}_{i,j}(\mathbf{r}) = \langle u_i(\mathbf{x})u_j(\mathbf{x}+\mathbf{r}) \rangle
\end{equation}
where $\mathbf{x}$ is an arbitrary point (which could be the MP described above) and $\mathbf{r}$ is the three-dimensional displacement. The brackets indicate ensemble averaging over velocity field realizations. We notice here that the displacement $\mathbf{r}$ is the linear distance from the MP in the frozen flow field. The corresponding energy density tensor is given by
\begin{equation}\label{eq:29}
\mathbf{F}_{i,j}(\mathbf{k}) = FT \left \{ \mathbf{R}_{i,j}(\mathbf{r}) \right \},
\end{equation}
where $\mathbf{k}$ is the three-dimensional wave vector. As these three-dimensional quantities are hard to handle experimentally, the common practice is to define the i'th component of the one-dimensional spectra in a principal coordinate system,
\begin{equation}\label{eq:30}
F_{i,i}(\mathbf{k}) = FT \left \{ R_{i,i}(\mathbf{r}) \right \}.
\end{equation}
However, as is well known, these one-dimensional spectra are aliased in the sense that spectral components at an angle to the axes appear at lower spectral values~\cite{TennekesLumley1972}.

Of crucial importance to turbulence modelling is the so-called total turbulent kinetic energy spectrum, where all spectral components of a given wave vector magnitude are summed to provide the kinetic energy spectrum:
\begin{equation}\label{eq:31}
E(k) = \frac{1}{2} \sum_i \int_{k=|\mathbf{k}|} F_{i,i}(\mathbf{k})\,d\mathbf{k}.
\end{equation}
The question here is how our new spectra relate to these classical ones.

The classical spectra only have statistical meaning for a homogeneous velocity field measured at a given point in time. The results are found as ensemble averages over (infinitely) many realizations. We shall relax the condition of homogeneity of the velocity field to consider also fields with a non-isotropic angular velocity distribution in a spherical coordinate system, $P_u(\phi,\theta)$ (to include for example a field with a stationary local mean velocity), but still require homogeneity in the sense that $P_u(\phi,\theta)$ is the same throughout space along any homogeneous direction. We can then write the averaging process for the covariance as
\begin{align}\label{eq:32}
R_{i,j}(r) & = \left \langle u_i(\mathbf{x}) u_j(\mathbf{x}+\mathbf{r}) \right \rangle \nonumber\\
            &= \left \langle \int P_u(\phi,\theta)u_i(\mathbf{x},\phi,\theta) \, d\phi \, d\theta \right.  \nonumber\\
            & \left. \cdot\int P_u(\phi',\theta')u_j(\mathbf{x}+r,\phi',\theta') \, d\phi' \, d\theta' \right \rangle_r
\end{align}
where $(r,\phi,\theta)$ are spherical coordinates, the bracket $\langle \, \rangle _r$ indicates ensemble average over realizations using the same scalar distance $r$, and $P_u(\phi,\theta)$ and $P_u(\phi',\theta')$ are identical because of homogeneity but with independently fluctuating direction angles.

$R_{i,j}(r)$ is now a one-dimensional statistical quantity, which is only a function of the distance $r$ from the measurement point. The total turbulent kinetic energy spectrum is
\begin{equation}\label{eq:33}
E(k) = \frac{1}{2} \sum_i \int e^{-i2\pi kr}R_{i,i}(r)\,dr.
\end{equation}
We shall now compare this to the new method with reference to Figure~\ref{fig:radiuscurvature} below.

\begin{figure}
\includegraphics[width=0.45\textwidth]{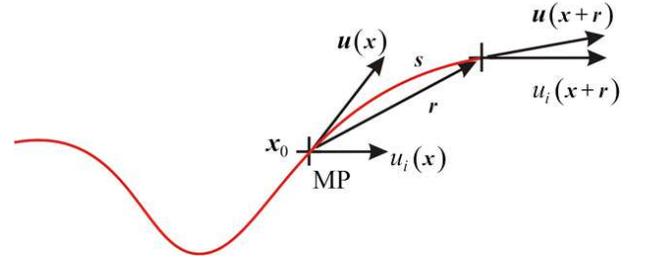}
\caption{\label{fig:radiuscurvature} Illustrating covariance measurements.}
\end{figure}

In the new situation, all measurements refer to the measurement point, MP, and the temporal record is converted to a spatial record consisting of a sum of convection elements. Since each convection element and the corresponding velocity vector are always co-parallel, we can write the one-dimensional covariance as a function of the spatial record, $s$:
\begin{align}\label{eq:34}
R(s) & = \left \langle u(s_0) u(s_0+s) \right \rangle \nonumber\\
            &= \left \langle \int P_u(\phi,\theta)u(s_0,\phi,\theta) \, d\phi \, d\theta  \right.  \nonumber\\
            & \left. \cdot \int P_u(\phi',\theta')u(s_0+s,\phi',\theta') \, d\phi' \, d\theta' \right \rangle_s
\end{align}
and the total turbulent kinetic energy spectrum as
\begin{equation}\label{eq:35}
E(k) = \frac{1}{2} \int e^{-i2\pi ks}R (s)\,ds.
\end{equation}
Note: This is a single term, one-dimensional expression, but since there are no velocity components normal to the convection, it includes the total turbulent kinetic energy. There is no need to add three terms. If we now compare the two paths in Figure~\ref{fig:radiuscurvature}, we can see that a certain distance $s$ along the convection record will not correspond to an equivalent distance $r$, unless the convection record can be considered approximately a straight line. In other words, the radius of curvature of the streak line should be large compared to the distance $s$.

Due to the discrepancy in path length between the accumulated arc length $s$ (eqn.~\ref{eq:1}) and the distance $r$, accurate direct comparison between the classical and new spectra can only be achieved within a limited spatial domain around the measurement point. The extent of this region may be formulated in terms of the radius of curvature, $R$, or the equivalent line curvature $\kappa = 1/R$. A measure classically associated with the spatial velocity curvature, or the gradients in the flow, is the Taylor microscale, $\lambda$. A relation between $\kappa$ and $\lambda$, based on scaling arguments, was deduced by Sch\"{a}fer~\cite{Schaefer2012,Schaefer2013}:
\begin{equation}\label{eq:36}
\left \langle \kappa ^2 \right \rangle \propto \lambda^{-2}
\end{equation}
or, in terms of radius of curvature:
\begin{equation}\label{eq:37}
R / \lambda = constant \quad \sim 1.
\end{equation}
The region within which the arc curvature can be considered negligible is therefore of the order of the Taylor microscale. As the Taylor microscale is an average obtained from flow statistics, so are the resulting curvature $\kappa$ and curvature radius $R$. Sch\"{a}fer~\cite{Schaefer2012,Schaefer2013} tested these results using DNS simulations of four different flows with different Reynolds numbers, displaying excellent agreement independent of Reynolds number within a range $Re_{\lambda} = 50 \leftrightarrow 300$.

One may therefore argue that, if one can assume local homogeneity and stationarity within this vicinity of the measuring point, there exists a direct correspondence between the classical and the here proposed energy spectra for the wavenumber range within the spatial extent of this region.

\section{\label{sec:LDAmeasurements}LDA measurements in a free jet in air}

We now apply our method to the measurement of a turbulent round jet in air. The free round turbulent jet is a canonical flow that, apart from the influence of the inlet, develops freely, unhampered by surrounding constraints such as walls. It is therefore an ideal experiment for the study of developing turbulence, both in the region near the jet exit, where the turbulence generation is  most active, and in the region further downstream, where the jet is fully developed. We have made measurements of the streamwise velocity component in radial scans from the jet center to points outside the jet interface to the surrounding air at distances from the jet exit of 10$D$, 15$D$ and 30$D$, where $D$ is the jet orifice diameter. The measurements were made with a side scattering laser Doppler anemometer (LDA) specially designed to obtain high resolution spectral information in a highly turbulent flow. The signal from the photodetector is digitized and saved, and all signal processing is performed in software.

The LDA with optical frequency shift and a spherical measurement volume is the only instrument that can make unbiased velocity measurement in the highly turbulent flow encountered between 10$D$ and 15$D$, especially in the shear layers and the outer region of the jet, and our method of converting time records into convection records allows us to study the convection of small spatial scales past the MP even if the flow is highly inhomogeneous (but stationary, allowing us to make sensible statistical calculations of e.g. energy spectra and structure functions).

The experiment is described in detail in Velte \textit{et al.}\cite{Velteetal2014a}. The jet data for present measurements were: Jet exit diameter: $D=10\, mm$, measurement location: 10, 15 and 30 diameters downstream at five points off axis (corresponding to 0, 0.5, 1.0, 1.5 and 2.0 jet half-widths, respectively). The jet exit velocity was $30\, ms^{-1}$, the measurement volume was $100\, \mu m$ in diameter and the average data rate was approx. $6400\, s^{-1}$ on the center line at downstream distance 30D from the nozzle exit. The scales are estimated as: Kolmogorov scale $= 53 \, \mu m$, Taylor scale $= 2.2\, mm$ at the center line at 30D downstream location. 4.000.000 data points were used at each location.

\begin{figure*}
\begin{minipage}[b]{0.45\linewidth}
\centering
\includegraphics[width=1.0\textwidth]{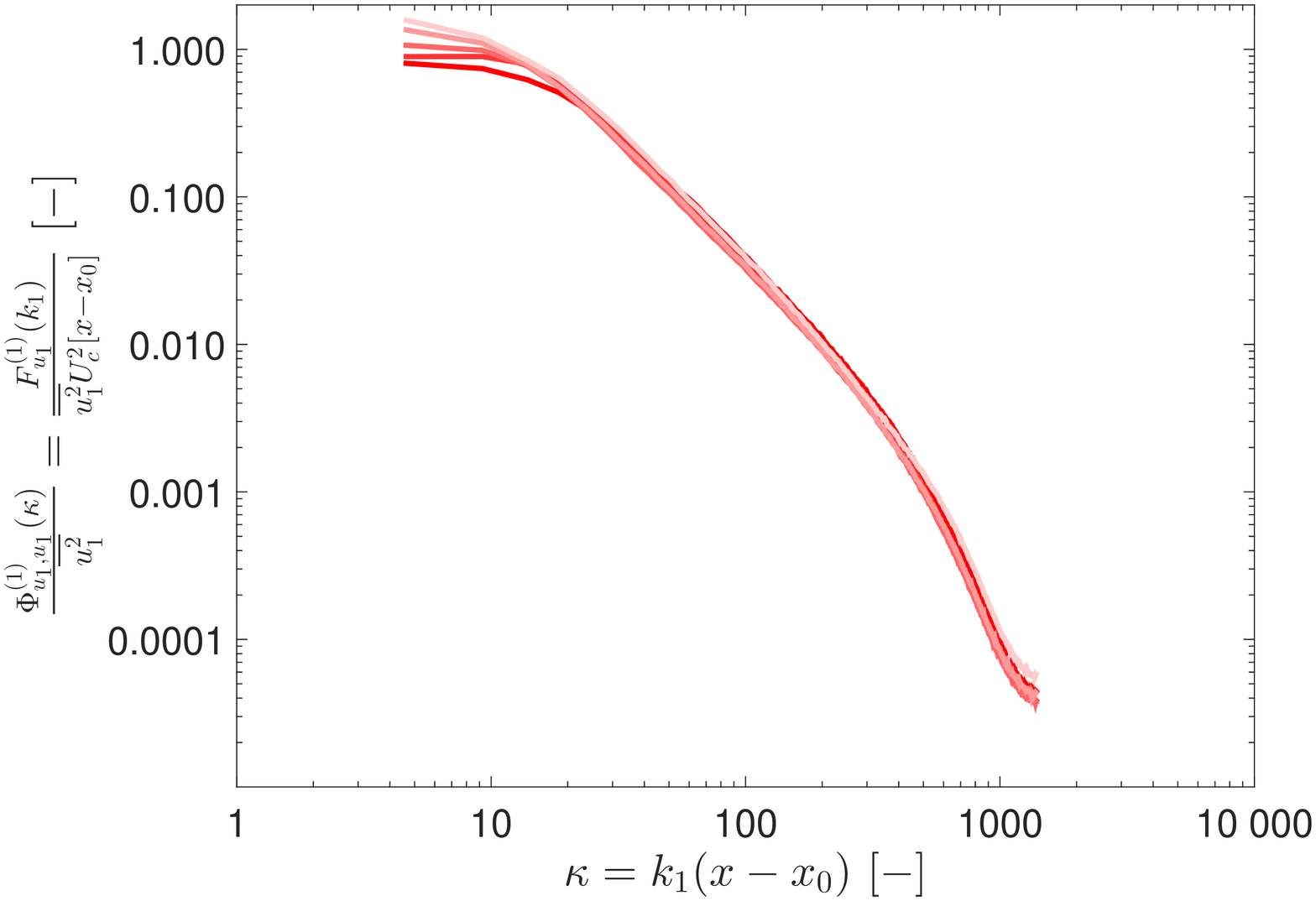}
\end{minipage}
\begin{minipage}[b]{0.45\linewidth}
\centering
\includegraphics[width=1.0\textwidth]{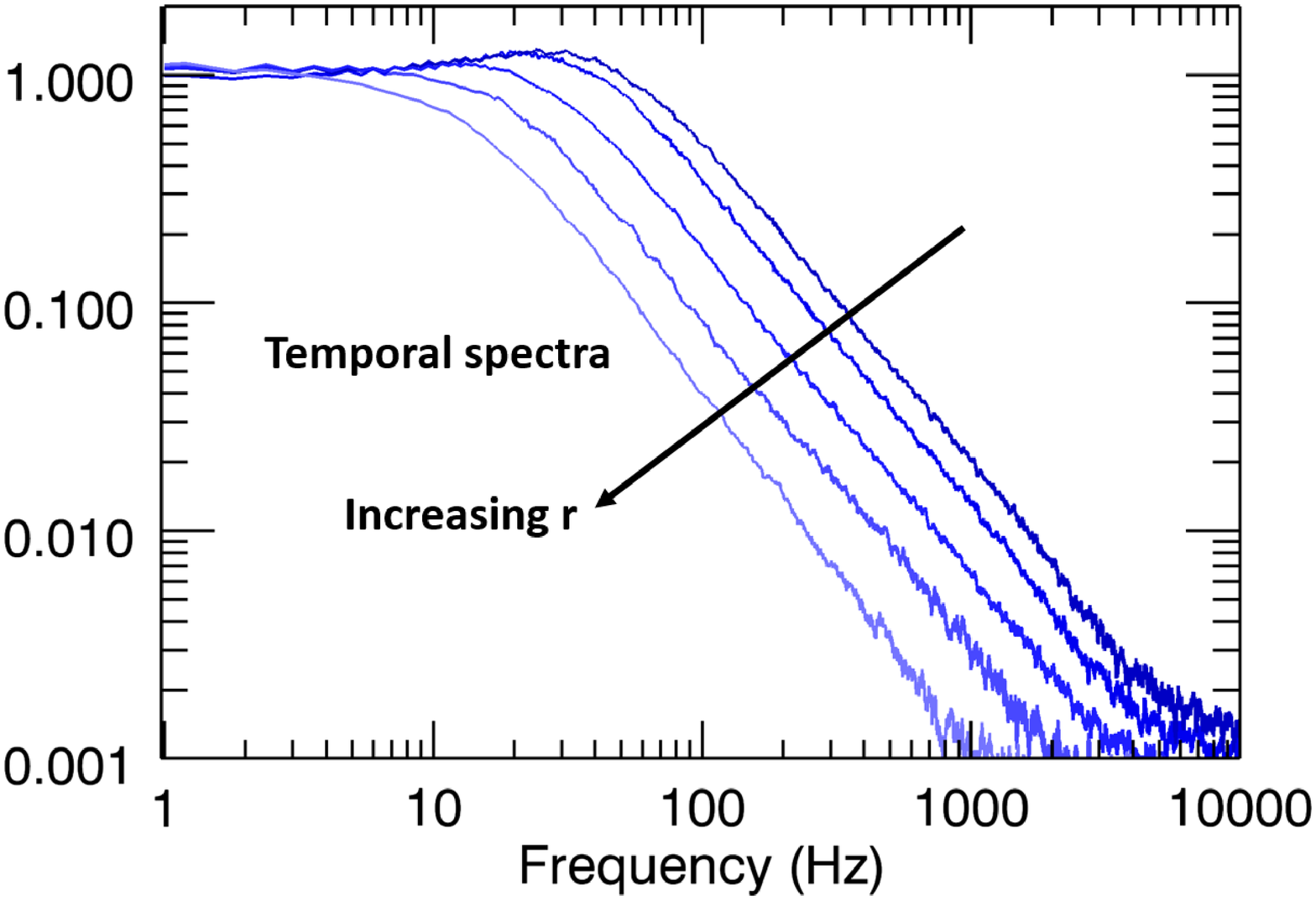}
\end{minipage}
\caption{\label{fig:Taylor-2_PR_fig4} LEFT: Spatial energy spectra measured using PIV along homogeneous directions in a mapped similarity space. The spectra have been normalized by each respective velocity variance. RIGHT: Corresponding temporal spectra of jet velocity at different off-axis positions (normalized). From heavy blue to light red and blue, respectively: off axis position $0,\,13,\,26,\,39,\,52\,mm$. Color online.}
\end{figure*}

\begin{figure*}
\begin{minipage}[b]{0.45\linewidth}
\centering
\includegraphics[width=1.0\textwidth]{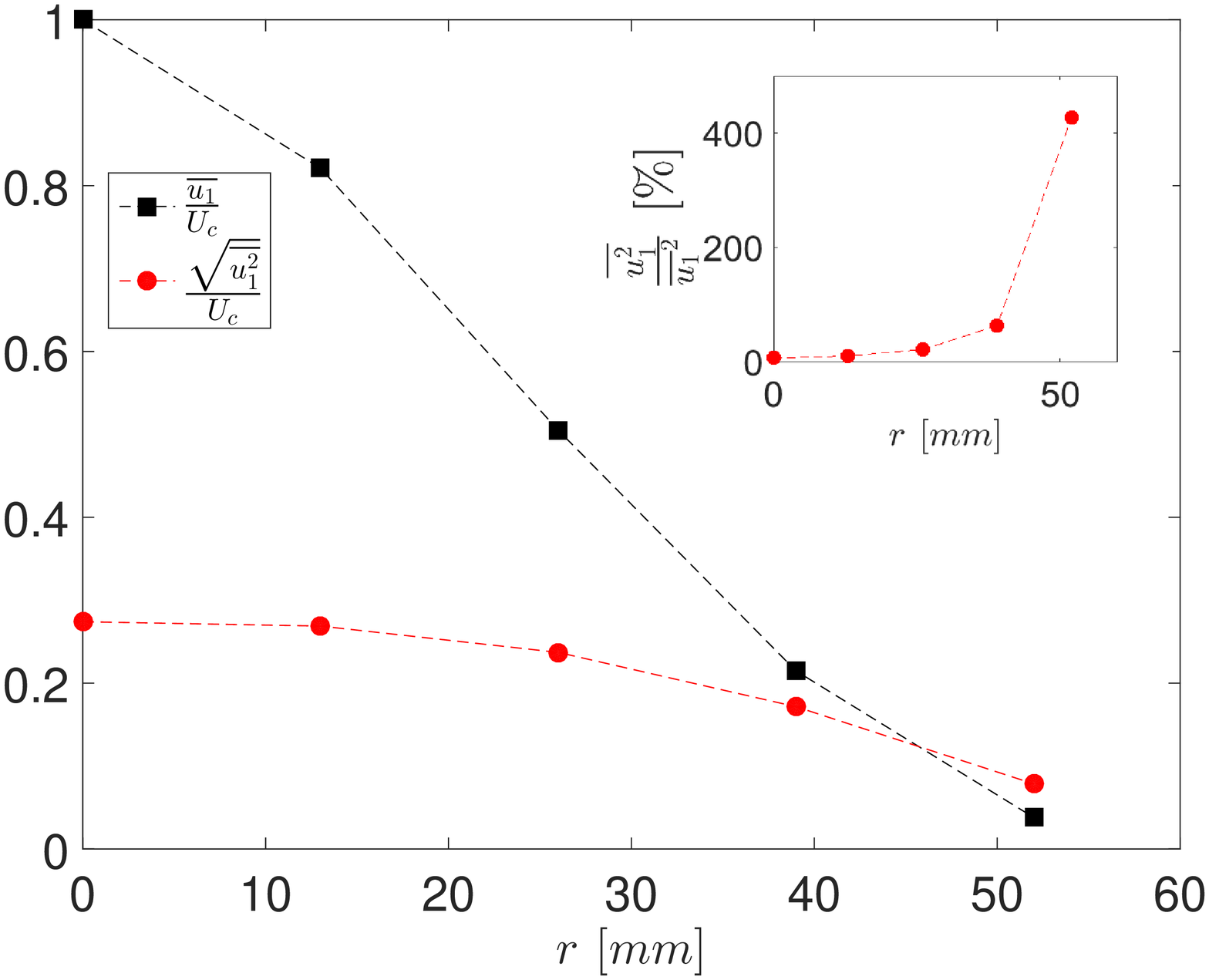}
\end{minipage}
\begin{minipage}[b]{0.45\linewidth}
\centering
\includegraphics[width=1.0\textwidth]{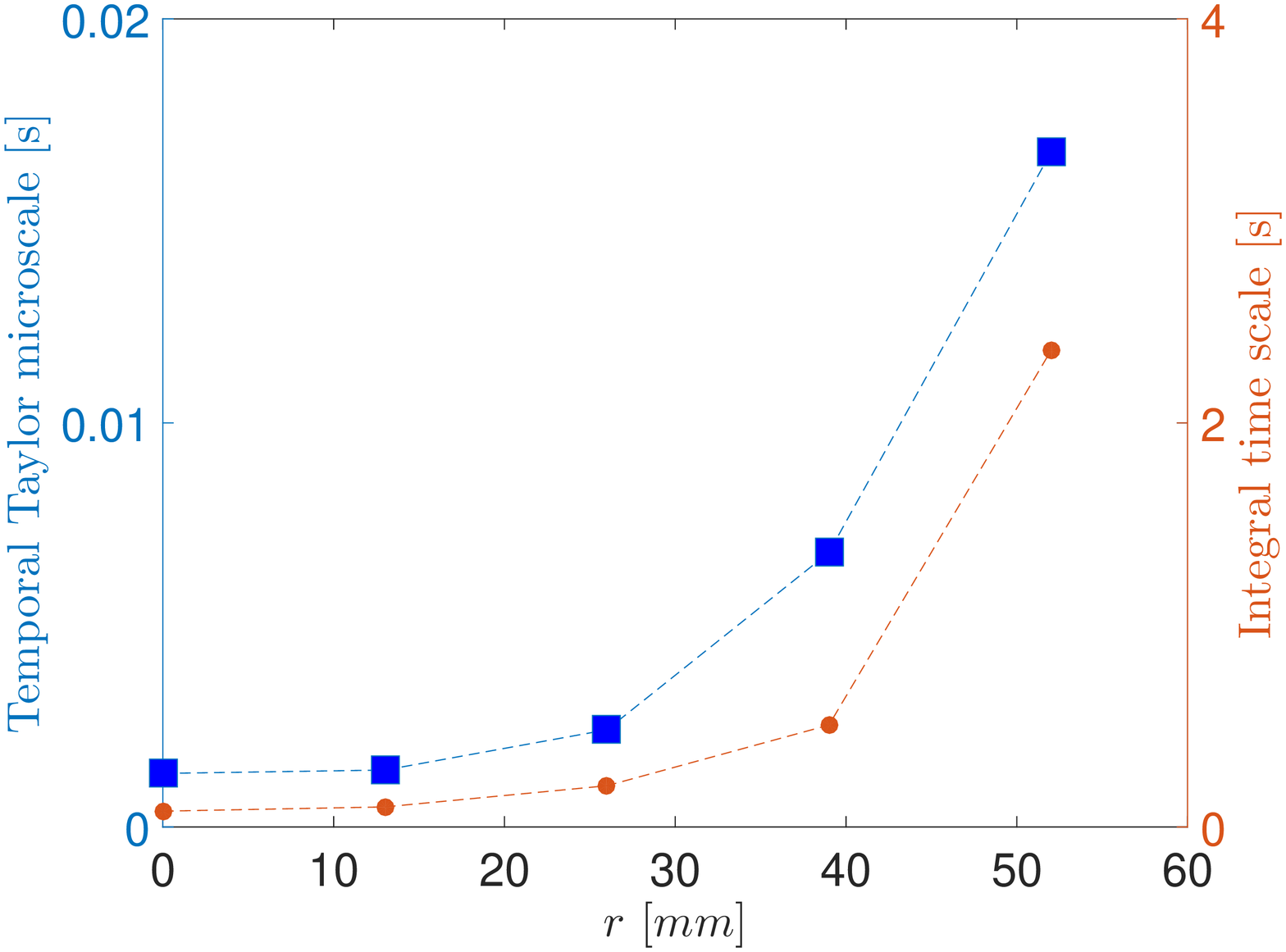}
\end{minipage}
\caption{\label{fig:values} LEFT: Measured mean velocity, velocity RMS and turbulence intensity at 30$D$ downstream of the jet nozzle at various radial distances from the jet centerline. RIGHT: Corresponding measured temporal Taylor microscales and integral timescales. Color online.} 
\end{figure*}

We first show results at 30$D$, where the jet is assumed to be fully developed and the turbulence is in equilibrium. As a baseline, spatial (non-dimensional) spectra measured by particle image velocimetry (PIV)~\cite{Hodzicetal2016} along homogeneous streamwise directions in a mapped similarity space~\cite{Ewingetal2007} are shown in Figure~\ref{fig:Taylor-2_PR_fig4} LHS. The spectra are shown to collapse when normalized by the local velocity variance, indicating that the spatial energy spectrum does indeed display the same energy distribution across scales, regardless of radial position in the jet. The temporal spectra measured by LDA are therefore normalized accordingly, see Figure~\ref{fig:Taylor-2_PR_fig4} RHS, resulting in normalized values at low frequencies which facilitates comparison of the shape of the spectra. As expected, the temporal energy spectra are shifted towards higher frequencies as the mean velocity increases. For completion, Figures~\ref{fig:values} displays the corresponding (LEFT) mean velocity, RMS velocity, turbulence intensity and (RIGHT) the temporal Taylor microscale and integral time scale measured with LDA for all radial positions measured. The Taylor microscale was extracted by taking the average time between zero crossings in the fluctuating velocity component time history. The integral time scale was extracted from the temporal energy spectrum extrapolated to zero frequency where the spectrum base level was adjusted from the dominating random noise level to a level where the integral under the one-sided spectrum matched half of the velocity variance, according to definition.

\begin{figure*}
\begin{minipage}[b]{0.45\linewidth}
\centering
\includegraphics[width=1.0\textwidth]{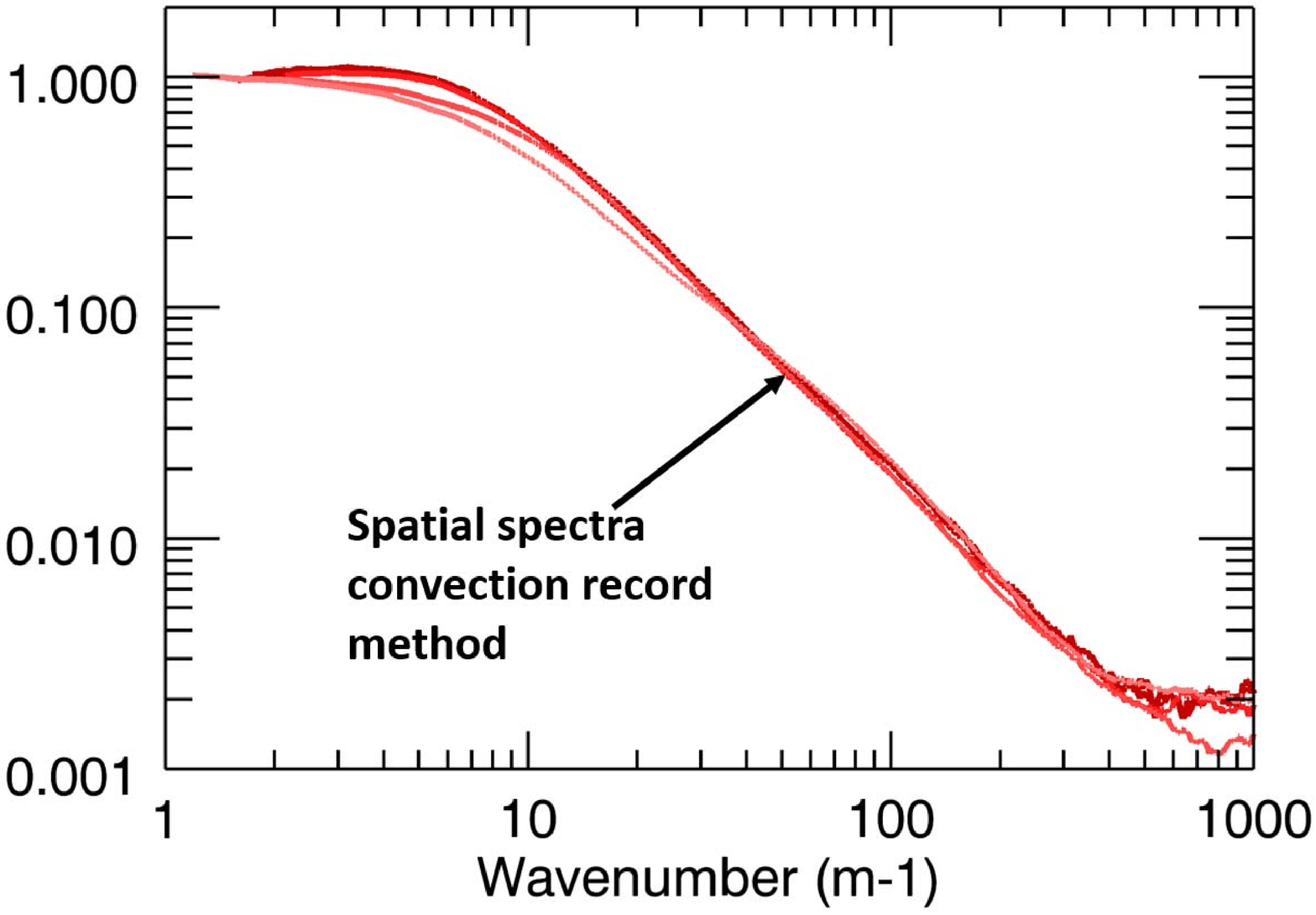}
\end{minipage}
\hspace{0.5cm}
\begin{minipage}[b]{0.45\linewidth}
\centering
\includegraphics[width=1.0\textwidth]{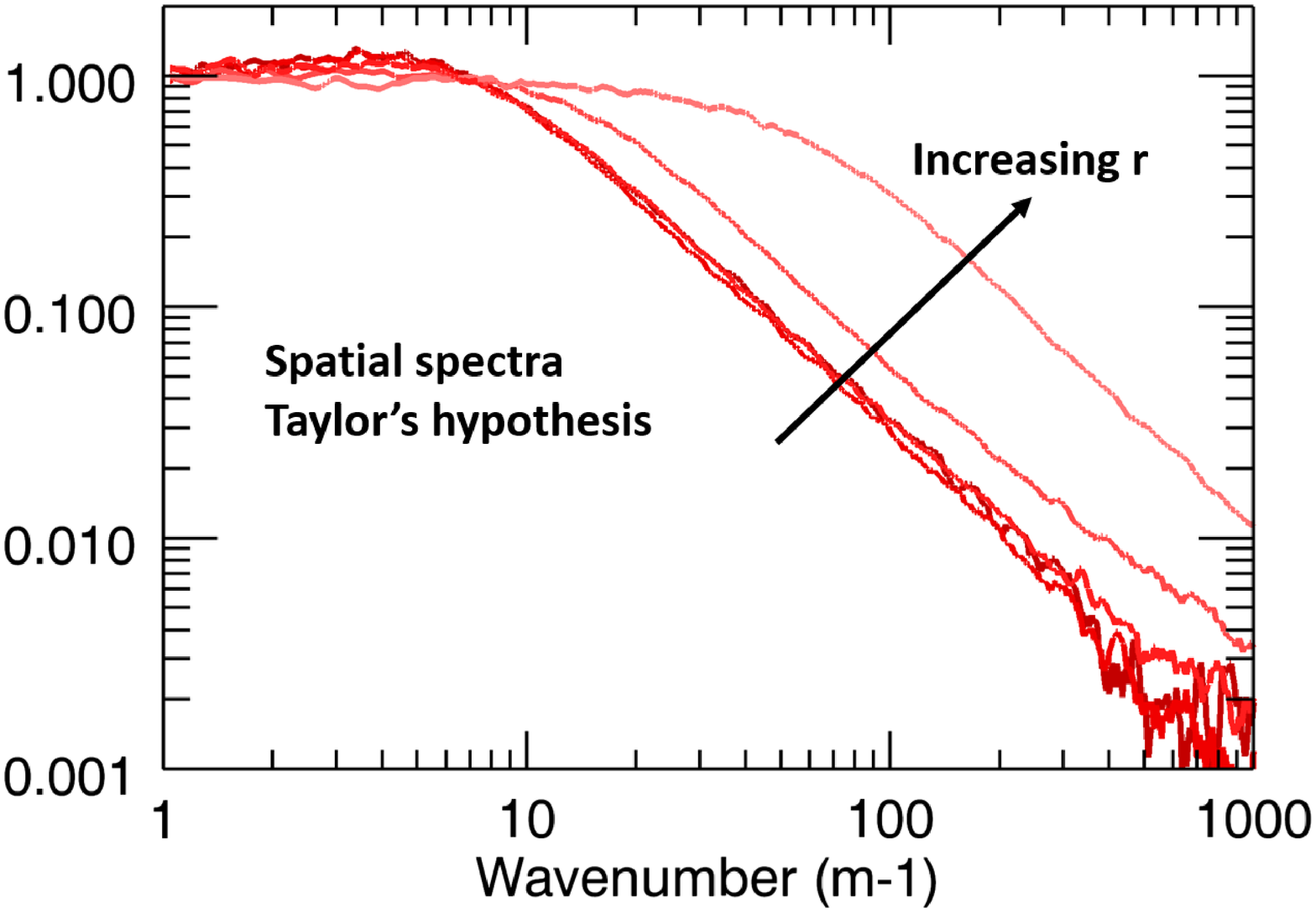}
\end{minipage}
\caption{\label{fig:Taylor-2_PR_fig5} LEFT: Restored spatial spectra using our method. Off-axis position heavy red to light red: 0, 13, 26, 39, 52$\,mm$. RIGHT: Taylor's hypothesis, same off-axis positions. Color online.} 
\end{figure*}

After application of our conversion method, we find the spatial spectra shown in LHS of Figure~\ref{fig:Taylor-2_PR_fig5}. The RHS shows spatial energy spectra based on Taylor's method. The spectra based on the new method collapse to nearly overlapping curves whereas Taylor's method deviates a little for the $26\, mm$ off-axis position and fails badly for the $39$ and $52\, mm$ off-axis positions. The turbulence intensity at these positions are 23\%, 65\% and 420\%, respectively. It is interesting to note that \textit{the collapse of the spatial spectra extend to low frequencies suggesting that the convection record may be valid also for scales larger than the Taylor microscale.}

We now proceed to the non-equilibrium part of the jet at 15$D$ and at 10$D$. Measurements in highly turbulent, non-equilibrium flows are difficult and prone to errors due to the fluctuating magnitude and direction of the instantaneous velocity. A series of measurement of velocity spatial energy spectra in non-equilibrium flows in wind tunnels behind regular and fractal grids are reported in~\cite{Laizet2015}. The spectra are examined with respect to the reference Kolmogorov spectrum and compared to an inertial subrange with a $-5/3$ slope. It was found, that even close to a fractal grid, a $-5/3$ slope could be detected over a wavenumber decade. As we show below, we find that the $-5/3$ slope we see at 30$D$ is reduced to just a tangential approximation to a $-5/3$ slope at 15$D$ and 10$D$. We believe that the $-5/3$ slope is to be expected in any flow with a uniform distribution of spatial scales as presumed in Kolmogorov's derivation.

\begin{figure*}
\begin{minipage}[b]{0.45\linewidth}
\centering
\includegraphics[width=1.0\textwidth]{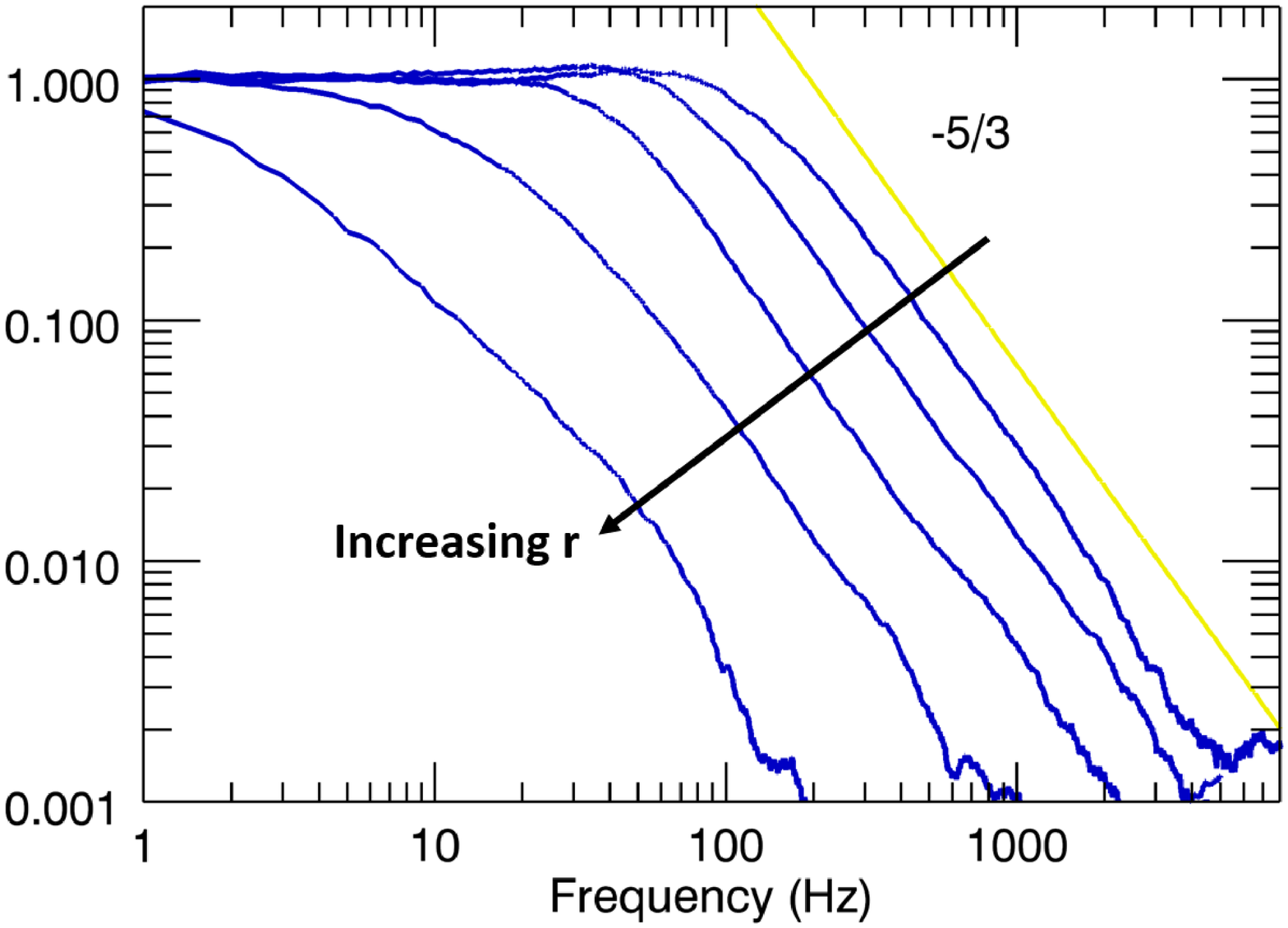}
\end{minipage}
\hspace{0.5cm}
\begin{minipage}[b]{0.45\linewidth}
\centering
\includegraphics[width=1.0\textwidth]{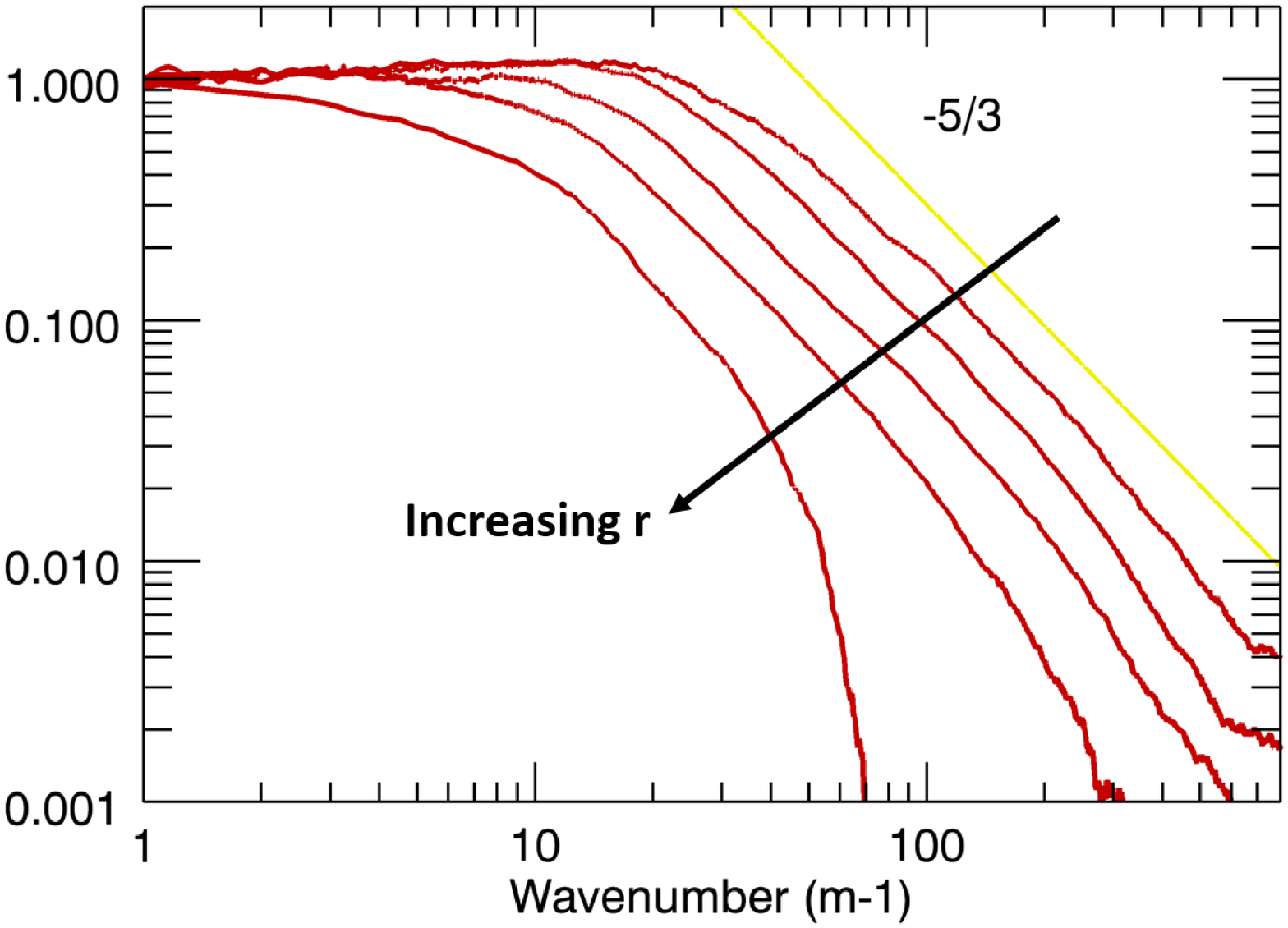}
\end{minipage}
\caption{\label{fig:15D} LEFT: Temporal energy spectra at 0, 0.5, 1.0, 1.5 and 2.0 jet half-widths at $x=15D$. RIGHT: Spatial energy spectra computed from the convection record at the same locations. Color online.}
\end{figure*}

\begin{figure*}
\begin{minipage}[b]{0.45\linewidth}
\centering
\includegraphics[width=1.0\textwidth]{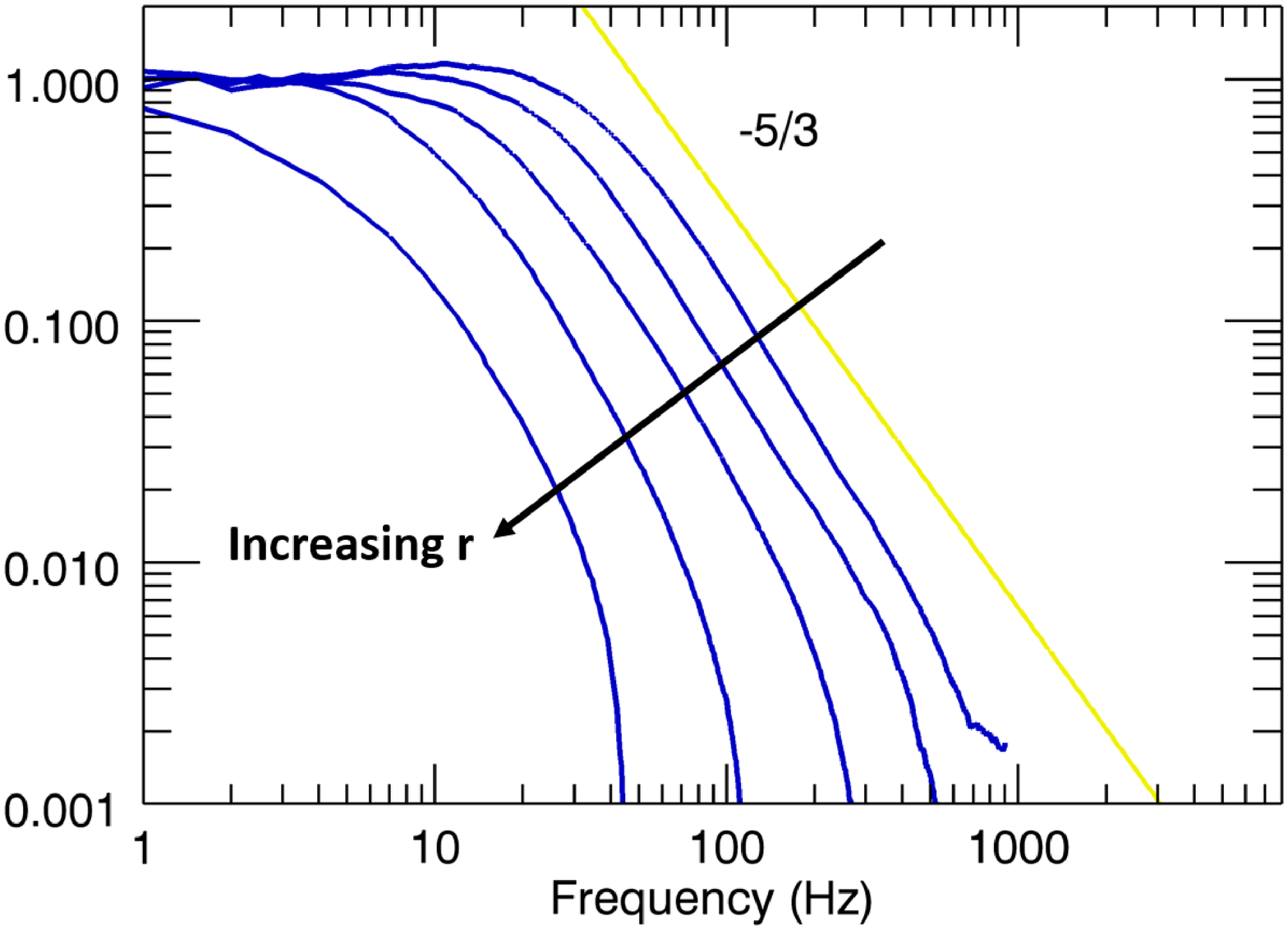}
\end{minipage}
\hspace{0.5cm}
\begin{minipage}[b]{0.45\linewidth}
\centering
\includegraphics[width=1.0\textwidth]{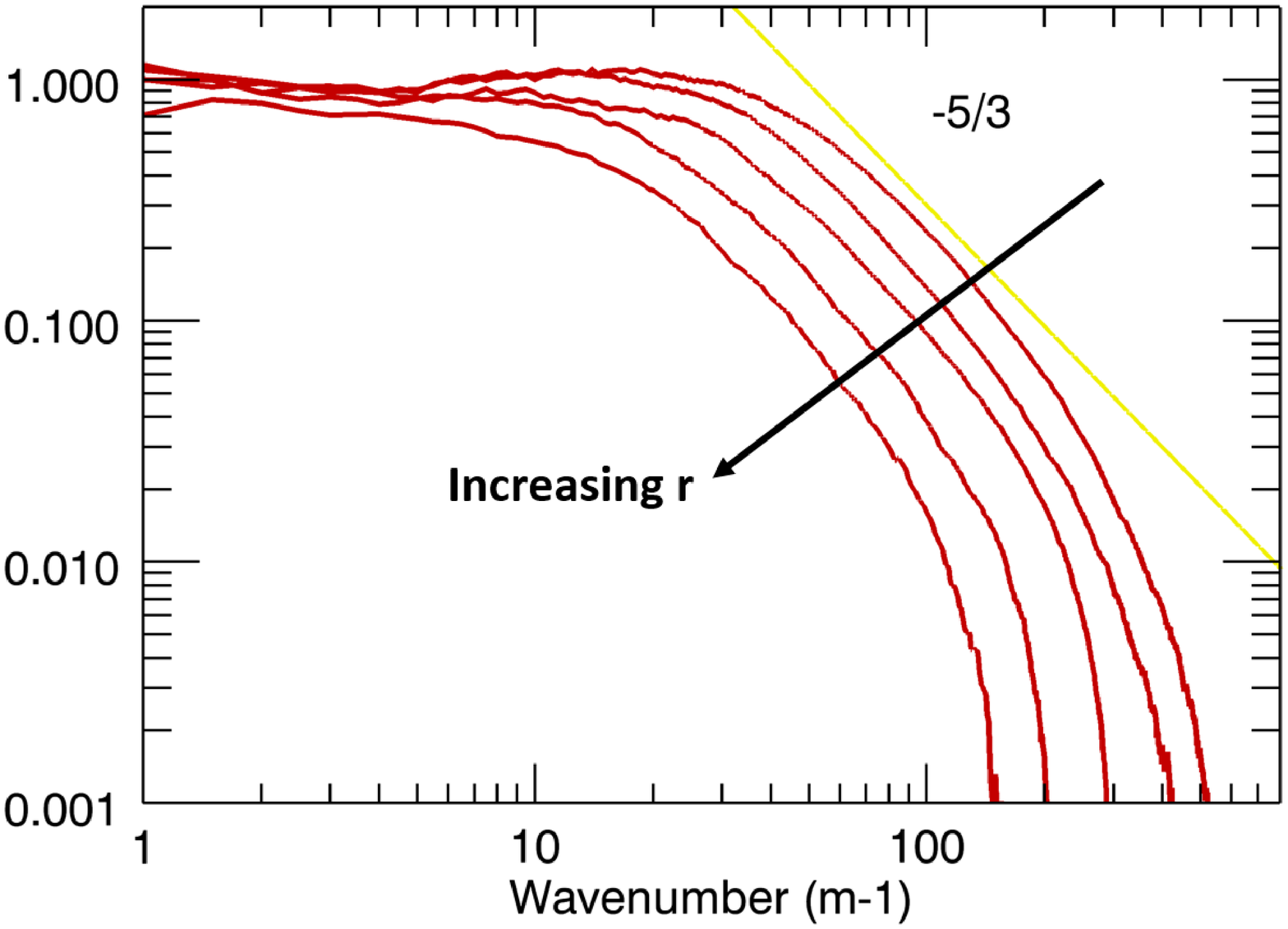}
\end{minipage}
\caption{\label{fig:10D} LEFT: Temporal energy spectra at 0, 0.5, 1.0, 1.5 and 2.0 jet half-widths at $x=10D$. RIGHT: Spatial energy spectra computed from the convection record at the same locations. Color online.} 
\end{figure*}

Figure~\ref{fig:15D} displays the temporal and spatial energy spectra measured at 15$D$ and Figure~\ref{fig:10D} displays the spectra at 10$D$. By inspection of these plots we may conclude the following.

\begin{itemize}
\item The temporal energy spectra are shifted to higher frequencies where the mean velocity is greater, as expected. However, unlike the situation at 30$D$, the spatial energy spectra do not collapse to a single curve indicating that the turbulence is not in equilibrium.
\item The slope of the high frequency / high wavenumber plots does not follow the $-5/3$ slope over any range, but do show a tangential approximation to the $-5/3$ slope. The match to the $-5/3$ slope is worst in the outer parts of the jet and at $x=10D$.
\item The deviation from the Kolmogorov spectrum appears to occur at the high frequency end of the spectrum, where the curves drop below the $-5/3$ slope. We hypothesize that the reason is that the Richarson-Kolmogorow cascade of large scales to small scales through triade interactions as described in e.g.~\cite{Kraichnan1971} has not had sufficient time to develop fully. Thus, the measurements give a rough indication of the dynamics of the cascade. The average convection velocity of the large eddies between the exit and the 10$D$ position is approximately $5\, ms^{-1}$. Thus, a convection time of $x / U_{conv} = 0.01\, m\, /\, 5\, m/s =0.01 / 5\, s = 20\, ms$ appears to be insufficient for the cascade to convert the large scales to the smallest scales that we can measure with our setup.
\item Finally, the different shapes of the curves near the axis and far from the axis seems to indicate that the turbulence is closer to the Kolmogorov shape near the axis than at the outer part of the jet.
\end{itemize}

\begin{figure*}
\begin{minipage}[b]{0.45\linewidth}
\centering
\includegraphics[width=1.0\textwidth]{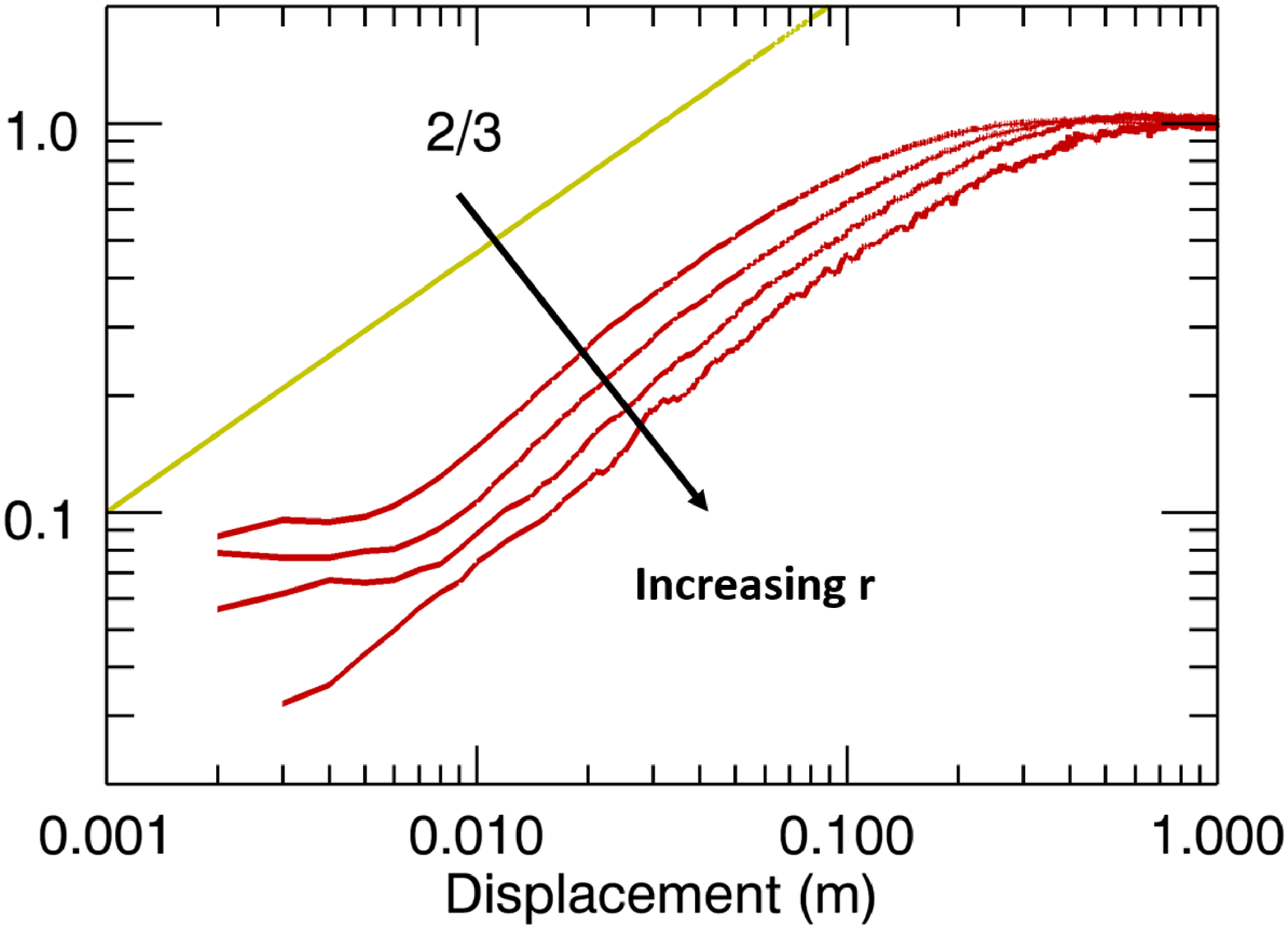}
\end{minipage}
\hspace{0.5cm}
\begin{minipage}[b]{0.45\linewidth}
\centering
\includegraphics[width=1.0\textwidth]{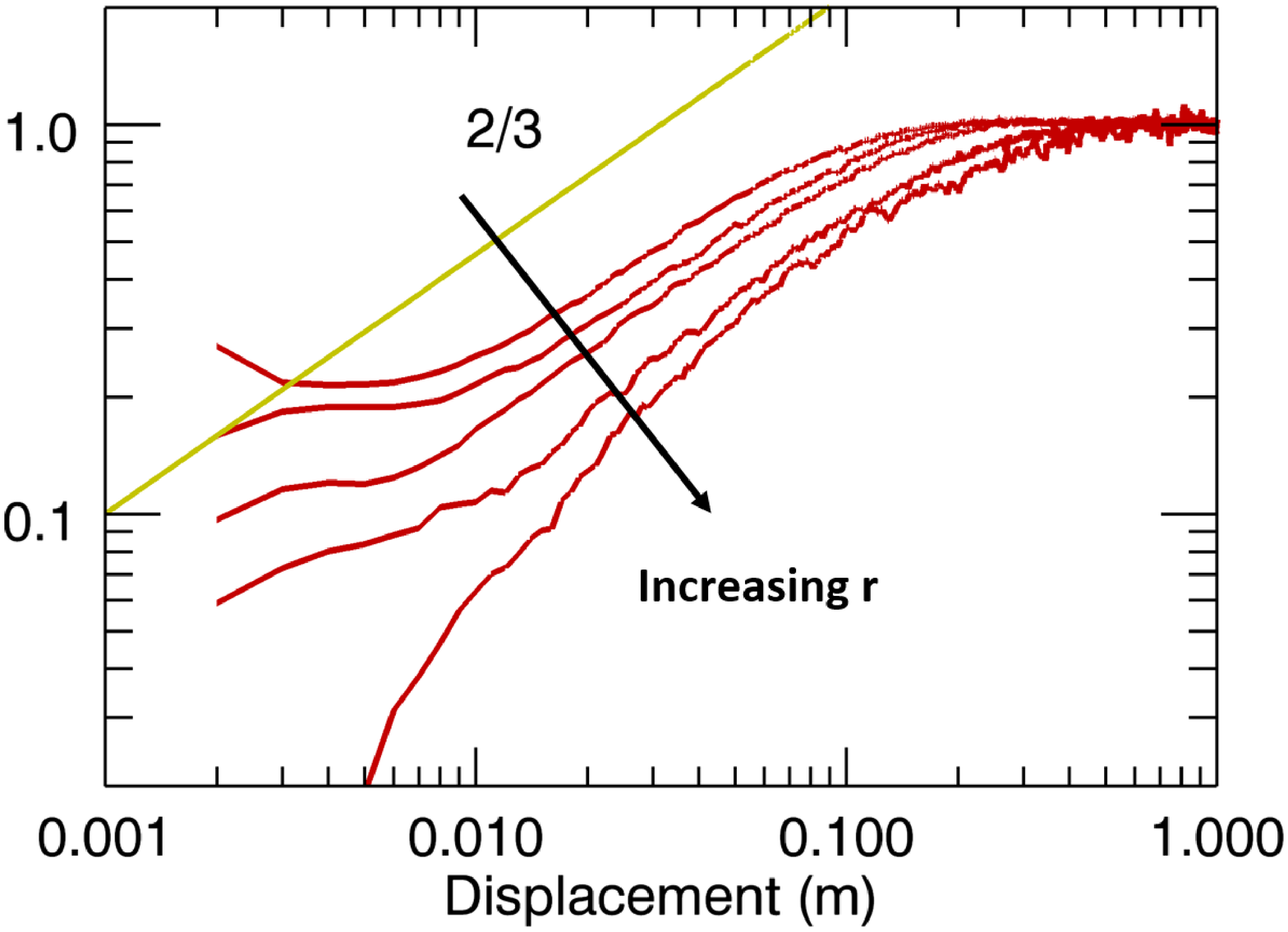}
\end{minipage}
\caption{\label{fig:struct} From left to right: 2nd order spatial structure functions at $15D$ and $10D$ and radial positions at 0, 0.5, 1.0, 1.5 and 2.0 jet half-widths. Color online.}
\end{figure*}

Figure~\ref{fig:struct} shows the measured data displayed as the 2nd order spatial structure function at the locations 15$D$ and 10$D$. We see again symptoms of non-equilibrium as differences in the curves at different radial positions. By inspection of the plots, we can see that there are relatively more large scale structures as we move downstream. We also notice that there is relatively more small scale activity near the center line than in the outer parts of the jet.

\section{\label{sec:Conclusion}Conclusion}

Our spatial record is not equivalent to a conventional streamline or streak line. It consists of the time-sampled velocities, but the temporal record is mapped into a spatial one-dimensional record consisting of a sum of consecutive convection  elements. We can interpret this record as describing the transport or convection of fluid parcels through the MV. With the fluid parcels follow fluid properties such as velocity structure, temperature, particle concentration etc.

It is intuitively clear that the scrambling of frequencies that occurs in the temporal energy spectrum because of the transport of small structures by the large instantaneous 3D velocity will be un-scrambled by the new method, at least as concerns the small, isotropic high wavenumber velocity structures. The spatial energy spectrum then expresses the energy of the spatial structures of the velocity passing the MV.

We show that spatial correlation functions and spatial energy spectra computed from the convection record are equivalent to the classical quantities within a spatial range defined by the Taylor microscale. Beyond this range, the statistics may still be useful though they are not necessarily directly comparable to their classical counterparts in a general flow setting. \textit{As can be observed in Figures~\ref{fig:Taylor-2_PR_fig4} LHS and~\ref{fig:Taylor-2_PR_fig5} LHS, comparing the spatial PIV spectra to the spatial LDA spectra, they do indeed produce directly comparable statistics across the complete spectrum of scales measured -- even well beyond the Taylor microscale.}

Our method completely bypasses the traditional Taylor's Hypothesis, but at the cost of an additional measurement of the magnitude of the instantaneous total 3D velocity vector. The method may be applied to all time sampling measurements (or flow simulations) obtained at a fixed point in (locally) stationary flow, but, in high intensity turbulent flow such as atmospheric and oceanographic flows, only the LDA (or maybe a 3D sonic anemometer) is able to provide reliable and unbiased high resolution velocity measurements.

It is important to note that this proposed convection record mapping is general and thus applicable independently of flow setting. Although the implementation is most straightforwardly and most accurately measured with correctly functioning laser Doppler anemometers, it can indeed be implemented on any set of regularly sampled data including computer simulations. The analysis is based on a continuous signal, so the discretized signal must reflect the same behavior by being sampled with a high enough sampling rate to resolve the smallest scales in the signal. If the smallest temporal/spatial scales of the analog measured signal cannot be resolved, the signal must be anti-aliasing filtered prior to sampling to avoid these effects according to standard theory~\cite{BendatPiersol}. The relevant parameter in LDA measurements is the mean sample rate of the randomly arriving temporal samples. To avoid averaging effects in the determination of the spatial sampling increment (Eq.~\ref{eq:24}), the average (and stationary) sample rate should be greater than the Nyquist rate.

We applied the method to measurements in a round, turbulent jet in air with radial scans at different downstream distances from the jet exit, both in the fully developed jet at 30$D$, where the turbulence is assumed to be in equilibrium, and in the strongly non-equilibrium parts of the jet at 15$D$ and 10$D$. In the fully developed part of the jet, the spatial spectra measured with LDA using our method show perfect agreement with spatial spectra derived from PIV measurements obtained in the same jet. Our spatial energy spectra and spatial 2nd order structure functions obtained in the non-equilibrium par of the jet reveal interesting features of the developing turbulence that can be interpreted as resulting from an incomplete Richardson -- Kolmogorov cascade process where the triadic interactions between large scales and small sales have not had time to reach equilibrium.

\begin{acknowledgments}
We wish to acknowledge the generous support of Fabriksejer, Civilingeni{\o}r Louis Dreyer Myhrwold og hustru Janne Myhrwolds Fond (grant journal no. 13-M7-0039 and 15-M7-0031) and Reinholdt W. Jorck og Hustrus Fond (grant journal no. 13-J9-0026). The authors also wish to thank Professor Emeritus Poul Scheel Larsen for many helpful discussions.
\end{acknowledgments}

\bibliography{pb_cmv_TH_r1}

\end{document}